\documentclass[
reprint,
superscriptaddress,
 amsmath,amssymb,
 aps,
prb,
floatfix,
]{revtex4-2}

\usepackage[export]{adjustbox}
 \usepackage{adjustbox}

\usepackage{dcolumn}% Align table columns on decimal point

\usepackage{amsmath}
\usepackage{physics}
\usepackage{svg}

\svgpath{{../imgs/}}
\usepackage{bm}% bold math
\usepackage{xcolor}
\begin{document}

\preprint{APS/123-QED}

\title{Dynamics and condensation of polaritons in an optical nanocavity coupled to two-dimensional materials}% Force line breaks with \\

\author{Maria Vittoria Gurrieri}
 \altaffiliation[]{mgurr@dtu.dk}%Lines break automatically or can be forced with \\
 \affiliation{%
Department of Photonics Engineering, Technical University of Denmark, 2800 Kgs. Lyngby, Denmark
}%

\affiliation{NanoPhoton - Center for Nanophotonics, Technical University of Denmark,
Ørsteds Plads 345A, DK-2800 Kgs. Lyngby, Denmark}
\author{Emil V. Denning}
\affiliation{%
Department of Photonics Engineering, Technical University of Denmark, 2800 Kgs. Lyngby, Denmark
}%
\affiliation{NanoPhoton - Center for Nanophotonics, Technical University of Denmark,
Ørsteds Plads 345A, DK-2800 Kgs. Lyngby, Denmark}

\affiliation{Nichtlineare Optik und Quantenelektronik, Institut f\"{u}r Theoretische Physik, Technische Universit\"{a}t Berlin, 10623 Berlin, Germany}

\author{Kristian Seegert}
\affiliation{%
Department of Photonics Engineering, Technical University of Denmark, 2800 Kgs. Lyngby, Denmark
}%
\affiliation{NanoPhoton - Center for Nanophotonics, Technical University of Denmark,
Ørsteds Plads 345A, DK-2800 Kgs. Lyngby, Denmark}

\author{Philip T. Kristensen}
\affiliation{%
Department of Photonics Engineering, Technical University of Denmark, 2800 Kgs. Lyngby, Denmark
}%

\affiliation{NanoPhoton - Center for Nanophotonics, Technical University of Denmark,
Ørsteds Plads 345A, DK-2800 Kgs. Lyngby, Denmark}
\author{Jesper Mørk}
\affiliation{%
Department of Photonics Engineering, Technical University of Denmark, 2800 Kgs. Lyngby, Denmark
}%

\affiliation{NanoPhoton - Center for Nanophotonics, Technical University of Denmark,
Ørsteds Plads 345A, DK-2800 Kgs. Lyngby, Denmark}

\date{\today}% It is always \today, today,

\begin{abstract}
We present a comprehensive investigation of the light-matter interaction dynamics in two-dimensional materials coupled with a spectrally isolated cavity mode in the strong coupling regime. The interaction between light and matter breaks the translational symmetry of excitons in the two-dimensional lattice and results in the emergence of a localized polariton state. Employing a novel approach involving transformation to exciton reaction coordinates, we derive a Markovian master equation to describe the formation of a macroscopic population in the localized polariton state. Our study shows that the construction of a large-scale polariton population is affected by correction terms addressing the breakdown of translational symmetry. Increasing the spatial width of the cavity mode increases the Coulomb scattering rates while the correction terms saturate and affect the system's dynamics progressively less. Tuning the lattice temperature can induce bistability and hysteresis with different origins than those recognized for quantum wells in larger microcavities. We identify a limit temperature $T_{\mathrm{l}}$ as a key factor for stimulated emissions and forming a macroscopic population, enriching our understanding of strong coupling dynamics in systems with extreme confinement. 
\end{abstract}

\maketitle

%\tableofcontents

\subsection{\label{sec:level0}Introduction}

In the study of light-matter interaction, strong coupling refers to the condition where the coupling strength between photons and excitations in a material dominates over dissipative processes \cite{coupling}. This gives rise to many fascinating phenomena because of the formation of polaritons, quasi-particles that combine the properties of light and matter. In particular, when the excitons in a quantum well or a two-dimensional (2D) semiconductor \cite{2d} hybridize with the light field confined in an optical cavity, a condensate of polaritons can be observed \cite{Deng2002,Kasprzak2006,Deng2007,AntonSolanas2021,Waldherr2018}. One mechanism behind the formation of a macroscopic population of quasi-particles has been traced back to stimulated scattering due to the quasi-bosonic nature of polaritons. Many experiments have been conducted to study 2D and zero-dimensional (0D) polariton systems \cite{Scafirimuto2017}, where these particles are confined in either one or three dimensions. Theoretical studies in quantum wells describe polariton dynamics through scattering processes conserving both energy and momentum \cite{tassone_mottdensity,Porras,phonon}. 
\begin{figure}[h!!]
\hspace*{-0.5cm}
\includegraphics[width=1\linewidth,left]{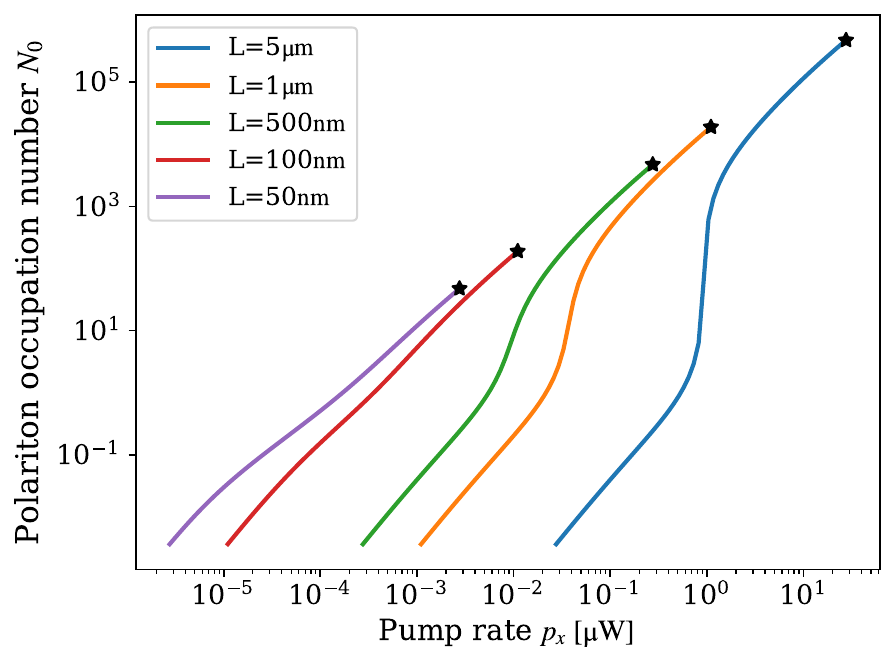}
\caption{Ground state polariton occupation as a function of pump power for different in-plane confinement lengths, $L$, of the optical cavity. The curves are computed by adapting the model from Ref.~\cite{Porras} to the case of a system with a single optical mode. For each confinement length, the star indicates the Mott transition.}
\label{fig:porras}
\end{figure}
These models predict that by increasing the confinement, i.e. reducing the in-plane cavity length, the threshold to quantum degeneracy effects is reached at progressively smaller pump powers. Such behaviour is illustrated in Fig.~\ref{fig:porras}, where the ground state polariton occupation number $N_0$ is plotted as a function of pump power $p_x$ for different confinement lengths of a 2D structure. The reduction in condensation threshold with in-plane confinement motivates the interest in studying polariton condensation in systems with extreme confinement. Current theoretical models, however, are derived for to 2D structures with transverse extent on the order of micrometres and larger. Therefore, the usefulness and validity of these models are not guaranteed when the optical cavity is scaled down to a transverse extent so small that the system response is dominated by a single or a few resonances rather than a continuum of modes. In particular, the assumption of translational symmetry used in previous works \cite{Porras} breaks down.

\begin{figure}[h!!]
\includegraphics[width=0.9\linewidth]{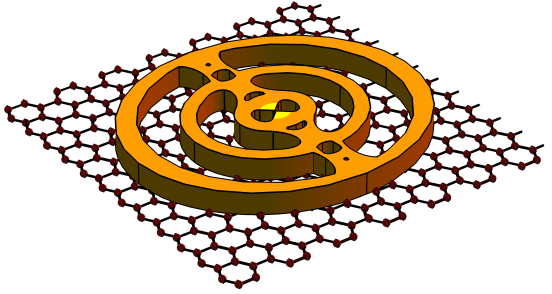}
\caption{Schematic representation of a sheet of 2D material coupled to an optical nanocavity, with a single optical mode localized at the centre of the bowtie. Strong light-matter coupling between the resonant electromagnetic field and the excitons in the 2D sheet leads to the formation of a localized polariton state.}
\label{fig:2dmaterial}
\end{figure}
In this work, we study light-matter interaction in a 2D semiconductor with a direct band gap placed on top of an optical nanocavity, as illustrated in Fig.~\ref{fig:2dmaterial}. Within this system, the 2D semiconductor excitations form a particle reservoir composed of thermalized excitons that can strongly couple to the resonant electromagnetic field of the nanocavity. The nanocavity design is taken from Ref.~\cite{george}, and we note that a similar cavity was recently experimentally demonstrated to implement deep sub-wavelength confinement \cite{nano}. Our goal is to reduce the threshold for detecting a macroscopic polariton population, providing, at the same time, an extensive reservoir that can pump particles into the condensate. The 2D structure coupled to the single cavity mode is modelled as an open quantum system in the framework of exciton reaction coordinates (ERC) \cite{ERC,ERC1,ERC2}, where the excited particles in the 2D material behave as interacting bosons. The dynamics of the polariton population are described by combining the formulation of a localized exciton state in the ERC formalism with the rate-equation approach previously developed by Tassone $et \ al.$ \cite{tassone_mottdensity} and Porras $et \ al.$ \cite{Porras}. The gain mechanism, which is induced by stimulated scattering, is responsible for a macroscopic population and the formation of a condensate in the lower polariton state. This condensate, in turn, is responsible for coherent light emissions and requires much lower excitation power than a standard laser. The reduced threshold for coherent emissions is attributed to the fact that condensation does not necessitate population inversion \cite{Imamoglu1996, Snoke2012}.

The article is organized as follows. In Sec.~\ref{sec:level1}, the theoretical model is set up. First, the ERC formalism is introduced. This enables a description in terms of a system where a single exciton is coupled to one mode of the electromagnetic resonator and a reservoir of residual excitons. The system's time evolution is described by a Markovian master equation from which a set of rate equations for macroscopic observables is derived. In Sec.~\ref{sec:level3}, the main results predicted by the theory are presented and explained. Specific attention is given to the understanding of how the temperature, the in-plane confinement, and the non-resonant pumping affect the dynamics. Sec.~\ref{sec:level5} presents the conclusions and further discussions.

\subsection{\label{sec:level1}Theoretical framework}
This section describes the mathematical tools employed to study the interaction between light in a general electromagnetic resonator dominated by a single resonance and excitons in an adjacent 2D system, as illustrated in Fig.~\ref{fig:2dmaterial} for the case of an optical nanocavity. In a semiconductor, the Coulomb interaction between electrons and holes results in the creation of excitons, which can freely move in a translationally invariant medium.
However, when introducing the electromagnetic resonator, the interaction between light and matter disrupts the translational invariance, giving rise to localized excitonic states with a centre-of-mass wave function exactly matching the electromagnetic field profile \cite{emil}. 
\begin{figure}[h!!]
\includegraphics[width=1\linewidth]{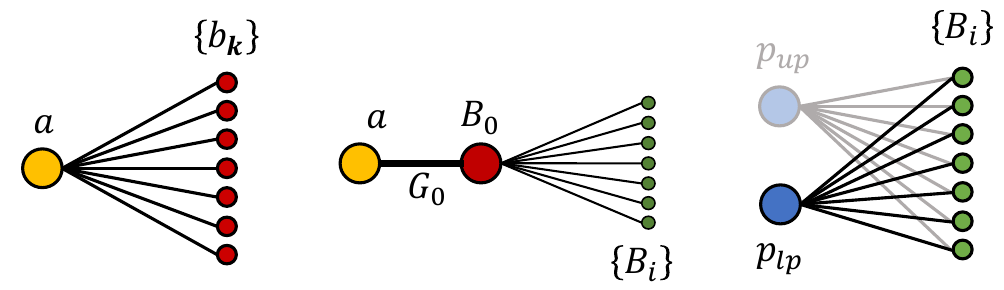}
 \caption{Schematic representation of the construction of the exciton reaction coordinate formulation. (a) The localized resonant field, with $a$ the operator describing the quantum properties of the resonant field, is coupled to a continuum of exciton modes with momentum $\mathbf{k}$, described by the annihilation operators ${b_{\mathbf{k}}}$. (b) The strength of the light–matter interaction is described through the coupling strength $G_0$ between the resonant field and a single collective exciton reaction coordinate with annihilation operator $B_0$, which, in turn, is coupled to an environment of residual exciton modes described by the operators $B_i$. (c) In the strong coupling regime, the Hopfield transformation \cite{Valatin1958, Bogoljubov1958} enables a description of the system through polariton quasi-particles with annihilation operators $p_{\mathrm{lp/up}}$, forming the upper and lower polariton branch. For the present applications, the former is energetically decoupled from the dynamics and is therefore neglected.}
\label{fig:exciton}
\end{figure}

If the light-matter coupling is sufficiently strong, it will give rise to a spatially localized polariton state as illustrated in Fig.~\ref{fig:exciton}. The system is initially described by coupling the resonant  electromagnetic field to all excitons with different in-plane momenta. In a change of basis, this is equivalent to the case where the electromagnetic field is coupled to only a single exciton reaction coordinate which, in turn, is coupled to a bath of residual exciton states. In the strong coupling regime, finally, a Hopfield transformation enables the identification of discrete lower and upper polariton states. As shown below, the population of the lower polariton branch can be modelled using a Born-Markov master equation, which considers the interaction with a continuous reservoir of excitons. This, in turn, enables a rate equation description for the dynamics of the lower polariton population. A complete description of the dynamics requires two additional equations that capture the evolution of the exciton density $n_x$, and the energy density $e_x$ in the surrounding 2D material. The scattering processes between the particles also modify the temperature and can lead to heating the surrounding lattice.

\subsubsection{Exciton reaction coordinate and strong coupling}
The model is based on Wannier-Mott exciton states and their interactions with a resonant electromagnetic field described by a single quasi-normal mode \cite{quasinorma1,quasinormal2,quasinormal3,quasinormal4}. For the reference calculations and the simulation in Sec.~\ref{sec:level5}, we consider a monolayer of transition metal dichalcogenide (TMDC). These materials present a direct band-gap at the $K$ and $K'$ points of the reciprocal lattice space \cite{Xiao2012,Cao2012}. Their advantage, compared to semiconductor quantum wells such as gallium arsenide, relates to the properties of monolayer TMDC excitons \cite{Li2014,Schneider2018,Lundt2016}. Their large binding energy makes them stable and enables the study of polariton condensation even close to room temperature \cite{AlAni2021}. Also, their small Bohr radius provides access to higher exciton densities before the Mott-transition is reached.

Wannier-Mott excitons are described via the relative position of the electron and hole and their centre-of-mass, which is free to move throughout the entire sheet of the 2D material. In practice, the centre of mass motion is described by a plane wave of momentum $\mathbf{k}$. Being composed of electron-hole pairs, excitons are represented by operators with neither bosonic nor fermionic commutation relations, which leads to exciton-exciton Coulomb interactions \cite{Rochat2000}. The excitations of the 2D material are, therefore, described as interacting bosons with raising and lowering operators $b^{\dagger}_{\mathbf{k}}$ and $b_{\mathbf{k}}$, respectively. We consider the case of an electromagnetic resonator with a single resonance in the bandwidth of interest for which the corresponding quantized electromagnetic field can be written in terms of the photon raising and lowering operators $a^{\dagger}$ and $a$, respectively. In the framework of the ERC, the interaction between excitons in the 2D material and the cavity mode is collected inside a single exciton with annihilation operator $B_0 = (\sqrt{\sum_{\mathbf{k}}{|g_{\mathbf{k}}|^2}})^{-1}\sum_{\mathbf{k}}g^*_{\mathbf{k}}b_{\mathbf{k}}$, where $g_{\mathbf{k}}$ is the coupling strength \cite{emil}. $B_0$ is, in turn, coupled to the exciton reservoir described by the operators $B_i =\sum_{\mathbf{k}}U_{i,\mathbf{k}}b_{\mathbf{k}}$, where $U_{i,\mathbf{k}}$ is the transformation matrix. The first row of the ERC transformation $U$, is $U_{0,\mathbf{k}} = (\sqrt{\sum_{\mathbf{k}}{|g_{\mathbf{k}}|^2}})^{-1}\sum_{\mathbf{k}}g^*_{\mathbf{k}}$ while the remaining ones are constructed via Gram-Schmidt orthogonalisation as orthonormal vectors such that $U$ is unitary. Furthermore, the part of the Hamiltonian governing the modes with $i>0$ is re-diagonalized. The power of the ERC thus relies on the fact that the exciton reservoir does not interact directly with the cavity field. A formal introduction to the exciton reaction coordinate in this kind of system is beyond the scope of this article, and we refer to \cite{emil} for an in-depth discussion.

The Hamiltonian of the system describes the single isolated exciton and the quantized electromagnetic field and can be written
\begin{equation}\label{H_is}
H'_{\mathrm{s}} =\hbar
\Omega_0 B^{\dagger}_0B_0 +\hbar\omega a^{\dagger} a+ \hbar G_0(B^{\dagger}_0a+a^{\dagger}B_0),
\end{equation}
where $\hbar$ is the reduced Planck constant, and the first two terms are the free energy of the exciton reaction coordinate and the optical field, respectively. $G_0$ is the light-matter coupling strength, which, in practice, is obtained by integrating the optical mode distribution as detailed in Ref.~\cite{emil}. 

In this work, we are particularly interested in the so-called strong coupling regime, where the coupling strength $G_0$ is so large that the exchange rate between $B_0$ and $a$ is faster than the rate of any competing dissipative process. In this regime, when the cavity energy is tuned across the exciton energy, the dispersion relations for light and matter anticross, which signifies the generation of two new states, known as the lower and the upper polariton branches. These are quasi-particles made of both light and matter. In the present model, the electromagnetic field is described by a single optical mode inside the cavity, which implies that only a single upper and lower polariton state is created. This is very different from the case of a quantum well embedded between Bragg mirrors, where continuous polariton branches arise \cite{Porras}.

To highlight the formation of the polariton quasiparticles, one can perform the Hopfield-Bogoliubov transformation \cite{Valatin1958, Bogoljubov1958} to recover the normal modes of the Hamiltonian described by Eq.~\eqref{H_is}. These are 
\begin{equation}\label{coefficient}
\begin{split}
    &p_{\mathrm{lp}}= X_{\mathrm{lp}}B_0 - C^*_{\mathrm{lp}}a\\
    &p_{\mathrm{up}}= C_{\mathrm{lp}}B_0 + X^*_{\mathrm{lp}}a,
\end{split}
\end{equation}
where $X_{\mathrm{lp}}$ and $C^*_{\mathrm{lp}}$ are the excitonic and photonic weights of the new quasi-particles. Equation \eqref{H_is} can thus be written in diagonal form
\begin{equation}\label{H_p}
    H'_{\mathrm{s}} =  \epsilon_{\mathrm{lp}} p^{\dagger}_{\mathrm{lp}}p_{\mathrm{lp}}+\epsilon_{\mathrm{up}} p^{\dagger}_{\mathrm{up}}p_{\mathrm{up}},
\end{equation}
where $p^{\dagger}_{\mathrm{lp/up}}$ and $p_{\mathrm{lp/up}}$ are the raising and lowering operators of the lower and upper polariton state, and $\epsilon_{\mathrm{lp/up}}$ are the corresponding energies.

In order to account for the fact that excitons and polaritons are interacting bosons, we add the leading interaction term describing Coulomb scattering to the Hamiltonian in Eq.~\eqref{H_is}. This accounts for self-interaction within the polariton state and reads \cite{emil}
\begin{equation}\label{coulomb1}
\hat{W}_0 =W'_0 |X_{\mathrm{lp}}|^4p^{\dagger}_{\mathrm{lp}}p^{\dagger}_{\mathrm{lp}}p_{\mathrm{lp}}p_{\mathrm{lp}},
\end{equation}
with the effective non-linear interaction strength
\begin{equation}\footnotesize
W'_0 =\sum_{\mathbf{k},\mathbf{k}',\mathbf{q}} W_{\mathbf{k},\mathbf{k}',\mathbf{q}}U_{0,\mathbf{k}+\mathbf{q}}U_{0,\mathbf{k}'-\mathbf{q}}U^*_{0,\mathbf{k}'}U^*_{0,\mathbf{k}}.
\end{equation}
From now on, the upper polariton branch will be neglected because it resides high in energy and does not affect the dynamics described here. The lower polariton will then be labelled by $p_0$ instead of $p_{\mathrm{lp}}$ to highlight the relation with the ERC-operator $B_0$. See Eq.~\eqref{coefficient}.

The Hamiltonian for the residual excitons $B_i$ forming the reservoir is
\begin{equation}\label{reservoir}
     H_{\mathrm{R}} = \hbar \sum_{i>0} \Omega_i B^{\dagger}_i B_i.
\end{equation}
The Coulomb interaction of the excitons is neglected here. This follows from the assumption that the only region of the exciton Hilbert space, where the exciton density is sufficiently large to give a significant contribution, is the exciton-exciton interactions within the ERC, as described by Eq.~\eqref{coulomb1}. When the characteristic confinement length of the electromagnetic field $L$, here between $20 \ \mathrm{nm}$ and $70 \ \mathrm{nm}$, is large compared to the exciton Bohr radius, $a_{\mathrm{B}} \approx 1.95\ \mathrm{nm}$, the transformation elements $U_{0,\mathbf{k}}$ decay on a momentum scale that is small compared to the momentum variation of $W_{\mathbf{k},\mathbf{k}',\mathbf{q}}$. In this regime it is reasonable to use the approximation $W_{\mathbf{k},\mathbf{k}',\mathbf{q}}\simeq W_{000}$ \cite{emil,carusotto}.

The polariton state interacts with the residual excitons of the environment through Coulomb scattering. We consider here the primary interaction in which two excitons in the reservoir scatter, generating a particle in the lower polariton state and a higher-energy exciton. The opposite process, where a polariton and a high-energy exciton generate two lower-energy excitons, is also taken into account. Fig.~\ref{fig:exciton} shows a schematic representation of such processes between the continuum exciton reservoir and the single isolated polariton state. The expression for $H_{\mathrm{int}}$ then reads
\begin{equation}\label{interaction}
H_{\mathrm{int}}=\sum_{j,i',j'}(\mathcal{U}_{j,i',j'}X_{\mathrm{lp}}p^{\dagger}_{0}B^{\dagger}_jB_{i'}B_{j'}+\mathrm{h.c.}),
\end{equation}
in which $\mathrm{h.c.}$ denotes the hermitian conjugate, and 
\begin{equation}\label{coulomb}
\footnotesize
\mathcal{U}_{j,i',j'}=\sum_{\mathbf{k},\mathbf{k}',\mathbf{q}}W_{000}U_{0,\mathbf{k}+\mathbf{q}}U_{j,\mathbf{k}'-\mathbf{q}}U^*_{i',\mathbf{k}'}U^*_{i,\mathbf{k}}.
\end{equation}
It follows that $W_{000}$ can be removed from the summation since it loses its dependence on the exciton in-plane momentum. This approximation will be beneficial later on.
\begin{figure}[h!!]
\hspace*{-0.5cm} 
\includegraphics[width=1.\linewidth]{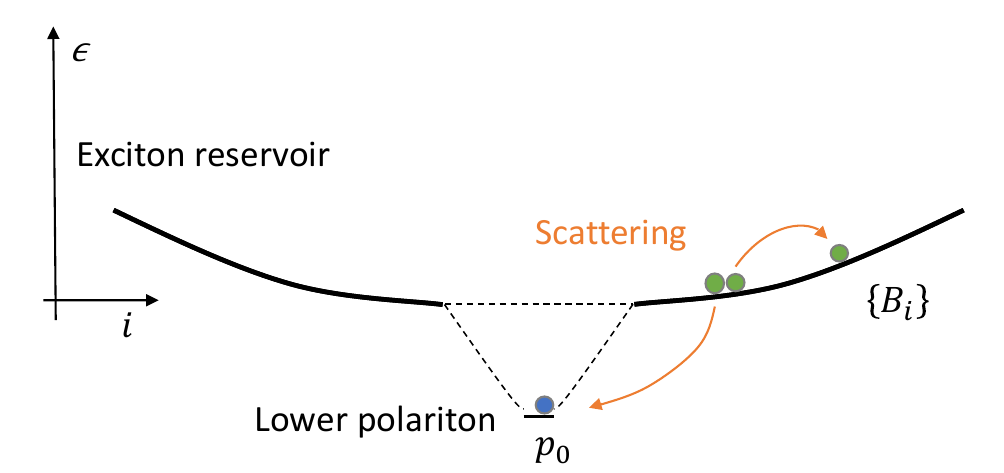}
\caption{Sketch of the scattering between two excitons in the reservoir. After the scattering, a particle is captured into the lower polariton state, and an exciton is scattered to a high-energy state in the reservoir. The subscript $i$ labels the bosonic states in the ERC and replaces the in-plane momentum $\mathbf{k}$ in the new basis.}
\label{fig:scattering}
\end{figure}

\subsubsection{Born-Markov Master Equation}
In the formalism of open quantum systems, the 2D material coupled to the nanocavity can be modelled with the Hamiltonian
\begin{equation}
    H= H_{\mathrm{s}} +  H_{\mathrm{R}} + H_{\mathrm{int}}
\end{equation}
in which the first term describes the system ERC-mode Hamiltonian in the polariton basis, in which $H_{\mathrm{s}} = H'_{\mathrm{s}} + W_0$ as defined in Eqs. \eqref{H_p} and \eqref{coulomb1}; $H_{\mathrm{R}}$ is given by Eq.~\eqref{reservoir} and encodes the exciton reservoir not interacting with the cavity mode; lastly, $H_{\mathrm{int}}$ is the system-environment interaction term described by Eq.~\eqref{interaction}.

The second-order perturbative Born-Markov master equation in the Schr{\"o}edinger picture, obtained by tracing out the environment, is given by \cite{book} 
\begin{equation}\label{sistem}\footnotesize
    \frac{\partial  \rho_{\mathrm{s}}(t)}{\partial t}=- \frac{i}{\hbar}[H_{\mathrm{s}},  \rho_{\mathrm{s}}(t)] -\frac{1}{\hbar^2} \int^{\infty}_{0}\mathrm{d}\tau  \mathrm{Tr_{\mathrm{R}}}[H_{\mathrm{int}},[\tilde{H}_{\mathrm{int}}(-\tau), \chi(t)]],
\end{equation}
where $\rho_{\mathrm{s}}$ is the system density matrix. The integrand is traced over the reservoir, and $\tilde{H}_{\mathrm{int}}(-\tau)$ is a time-dependent Hamiltonian operator. Eq.~\eqref{sistem} has the explicit expression
\begin{equation}\label{rate_derivation}\footnotesize
\begin{split}
    \frac{\partial  \rho_{\mathrm{s}}(t)}{\partial t}= & -\frac{\mathrm{i}}{\hbar}[H_{\mathrm{s}},  \rho_{\mathrm{s}}(t)]+\frac{1}{\hbar^2} \int^{\infty}_{0}\mathrm{d}\tau\Big((p'_0 \rho_{\mathrm{s}}(t)p^{\dagger}_0-p^{\dagger}_0p'_0  \rho_{\mathrm{s}}(t))\left<F^{\dagger}_0F'_0\right>\\
    &+(p'^{\dagger}_0 \rho_{\mathrm{s}}(t)p_0-p_0p'^{\dagger}_0 \rho_{\mathrm{s}}(t))\left<F_0 F_0^{'\dagger} \right>-\mathrm{h.c.}\Big),
        \end{split}
\end{equation}
where the operators labelled with prime are still in the interaction picture while
\begin{equation}\label{six}\footnotesize
F^{\dagger}_0=\sum_{ijh}\mathcal{U}_{i,j,h}B^{\dagger}_iB_jB_h
\end{equation}
describes the scattered excitons, Eq.~\eqref{coulomb}.

\subsubsection{Rate equations}
To study the dynamics of the lower polariton population $N_0=\left<p^{\dagger}_0p_0\right>$, we follow Ref.~\cite{Porras_2003} to develop a set of rate equations. We generalize this model to take into account the breaking of the in-plane translational symmetry induced by the nanocavity \cite{emil}. The evolution of the exciton density $n_x$ and the reservoir temperature $T_x$ are also described, leading to a set of self-consistent rate equations. 

The time evolution of the expectation value of the lower polariton population is given as {\footnotesize{$\dfrac{\mathrm{d}}{\mathrm{d}t}\left<p^{\dagger}_0p_0\right>= \mathrm{Tr}\Big(p^{\dagger}_0p_0\dfrac{\partial \rho_{\mathrm{s}}}{\partial_t}\Big)$}}, which leads to the equation
\begin{equation}\label{rate1}  \footnotesize
\begin{split}
    \frac{dN_0}{\mathrm{d}t}=&-\frac{1}{\hbar^2} \int^{\infty}_{0}\mathrm{d}\tau
        \Big( \left<p^{\dagger}_0p_0p^{\dagger}_0p'_0\right> - \left<p^{\dagger}_0p^{\dagger}_0p_0p'_0\right> \Big) \left<F^{\dagger}_0F'_0\right>\\
        &+\Big( \left<p^{\dagger}_0p_0p_0p^{'\dagger}_0\right>-\left<p_0p^{\dagger}_0p_0p^{'\dagger}_0\right> \Big)\left<F_0F^{'\dagger}_0 \right> + \mathrm{h.c.}.
\end{split}
\end{equation}
The multiple operator expectation values are expanded using Wick's theorem \cite{Wick1950}. For the six operators in $\left<F^{\dagger}_0F'_0\right>$ and $\left<F_0F^{'\dagger}_0\right>$, the assumption of a thermalized Maxwell-Boltzmann reservoir allows us to simplify the computation. We make the assumption that the electromagnetic mode function is both separable and localized. In this approximation, the precise overlap of the exciton reaction coordinate and the resonant field results in a coupling strength $G_0$ that remains unaffected by the electromagnetic field's lateral extent \cite{emil}. Furthermore, the lateral field distribution of the electromagnetic field is taken as Gaussian, which simplifies the computation of the Coulomb in-scattering rate and the out-scattering rate of the lower polariton state. The full derivation is reported in Appendix \ref{App_A}. The final expression for the time-dependence of $N_0$ then reads
\begin{equation}\label{rate5}
\begin{split}
    \frac{dN_0}{dt}=& (N_0+1)n^2_xW_{\mathrm{in}} - N_0n_x W_{\mathrm{out}_\mathrm{1}}-N_0n^2_x W_{\mathrm{out}_\mathrm{2}}\\
    &+[n^3_xW_{\mathrm{in_c}}-n^2_xN_0W_{\mathrm{out_c}}]-  \Gamma_{\mathrm{lp}} N_0,
    \end{split}
\end{equation}
where $W_{\mathrm{in}}$ and $W_{\mathrm{in}_\mathrm{c}}$ are the two rates of scattering into the lower polariton, scaling as $n_x^2$ and $n_x^3$, respectively. $W_{\mathrm{out}_\mathrm{1}}$,  $W_{\mathrm{out}_\mathrm{2}}$ and $W_{\mathrm{out}_\mathrm{c}}$ are the out-scattering rates, the former rate scaling with $n_x$  and the others with $n_x^2$. In the following, we use the label $W$ to refer to all the scattering rates.  Crucially, all the $W$-coefficients depend on the exciton temperature. The rates are thus not fixed quantities but are time-dependent. The last term of Eq.~\eqref{rate5} has been added phenomenologically to account for the radiative losses of the polaritons. In particular, $\Gamma_{\mathrm{lp}}= C^2_{\mathrm{lp}}/\tau_{\mathrm{ph}}+X^2_{\mathrm{lp}}/\tau_{x}$ where $\tau_{\mathrm{ph}}$ is the cavity photon lifetime and $\tau_{x}=\Gamma^{-1}_x$ is the exciton lifetime \cite{Deng2010}. Using the Maxwell-Boltzmann distribution, we define the exciton density $n_x$ through
\begin{equation}\label{occupation_number}
    N^x= \frac{2\pi n_x}{\rho_x k_{\mathrm{B}} T_x},
\end{equation}
in which $N^x$ is the total occupation number \footnote{the $i_{th}$ exciton state occupation number is $N^x_i=N^x\exp{-\beta \epsilon_i}$, with $\beta= (T_xk_{\mathrm{B}})^{-1}$ and $\epsilon_i$ the corresponding energy.}, $\rho_x=(m_{\mathrm{h}}+m_{\mathrm{e}})/\hbar^2$ is the 2D density of state that does not depend on space but only on the electron and hole masses $m_{\mathrm{e}}$ and $m_{\mathrm{h}}$, respectively. Finally, $k_{\mathrm{B}}$ is the Boltzmann constant, and $T_x$ is the exciton reservoir temperature, which evolves with the pumping as discussed below.

A self-consistent model for the dynamics requires the inclusion of two additional equations. These govern the changes in the density and energy of the exciton reservoir and are derived from Eq.~\eqref{rate5} by accounting for particles added to or removed from the system. The evolution of the exciton density involves a non-resonant pump to account for  the injection of thermalized excitons into the reservoir, $p_x$. Whenever a particle undergoes scattering into the lower polariton state, an exciton is extracted from the reservoir, and a second exciton with higher energy is produced, as depicted in Fig.~\ref{fig:scattering}. This process results in increasing the reservoir temperature. The third rate accounts for these dynamical changes in the exciton reservoir temperature. Consequently, the dynamics of $n_x$ can be described as follows
\begin{equation}\label{rate_2}
    \begin{split}\footnotesize
    \frac{dn_x}{dt}&=-(N_0+1)n^2_x\frac{W_{\mathrm{in}}}{S}+ N_0n_x \frac{W_{\mathrm{out}_\mathrm{1}}}{S}+\\
    &N_0n^2_x \frac{W_{\mathrm{out}_\mathrm{2}}}{S}+p_x-n_x\Gamma_x+N_0n^2_x W_{\mathrm{out}_\mathrm{2}}- \\
    &[n^3_xW_{\mathrm{in_c}}-n^2_xN_0W_{\mathrm{out_c}}]\frac{1}{S}
    \end{split}
\end{equation} 
where $S$ is the area of the 2D sheet. The non-resonant pump rate $p_x$ is modelled as a flux, having dimensions of inverse time per inverse square meter and scaling with the area of the 2D sheet.  In the absence of polaritons, the density of the excitons in the reservoir created by the pump is then $n_x = p_x \tau_x$. 
For $T_x = e_x/k_{\mathrm{B}}$
\begin{equation}\label{rate_3}
    \begin{split}\footnotesize
    \frac{de_x}{dt}&=-(N_0+1)n^2_x\frac{W_{\mathrm{in}}}{S}\epsilon_{\mathrm{lp}}+ N_0n_x \frac{W_{\mathrm{out}_\mathrm{1}}}{S}\epsilon_{\mathrm{lp}}+\\
    &N_0n^2_x \frac{W_{\mathrm{out}_\mathrm{2}}}{S}\epsilon_{\mathrm{lp}}+ p_xk_{\mathrm{B}}T_{\mathrm{L}}-n_x\Gamma_xk_{\mathrm{B}}T_x-\\
    &[n^3_xW_{\mathrm{in_c}}-n^2_xN_0W_{\mathrm{out_c}}]\frac{\epsilon_{\mathrm{lp}}}{S}
     \end{split}
\end{equation}
where $e_x$ is the energy density of the exciton reservoir, which evolves according to the equation, while $\epsilon_{\mathrm{lp}}$ is the energy of the single-state lower polariton branch. The term $p_xk_{\mathrm{B}}T_{\mathrm{L}}$, is considered here to be given by the injection of excitons at the lattice temperature $T_{\mathrm{L}}$. One should stress the difference between the lattice temperature, which accounts for the temperature at which the experiment is performed, and the exciton temperature, which is a dynamical variable. 

Aside from the inclusion of the second-order out-scattering rate, $W_{\mathrm{out}_\mathrm{2}}$, Eqs. \eqref{rate5}, \eqref{rate_2} and \eqref{rate_3} have two extra terms compared to previous models developed for quantum wells using the same methods \cite{Porras_2003}. Specifically, two additional scattering rates, $W_{\mathrm{in_c}}$ and $W_{\mathrm{out_c}}$, arise due to the loss of translation invariance in the system, see Appendix \ref{App_A}.  

\subsubsection{Scattering rates}\label{scattering}
Here, we discuss the physics of the Coulomb scattering rates entering Eqs. \eqref{rate5}, \eqref{rate_2} and \eqref{rate_3}. We also explain the impact of the loss of translation invariance on the system's dynamics. 

Due to the translational symmetry breaking the scattering rates do not have an analytic expression but must be computed numerically. We report here the expression for $W_{\mathrm{in}}$. Further details can be found in Appendix \ref{App_B} together with expressions for the other rates,
\begin{equation}\label{W_in_}
    \begin{split}
    W_{\mathrm{in}}=&\frac{4\hbar |W_{000}|^2|X_{\mathrm{lp}}|^2L^2 S^2 \rho_x}{2\pi (k_{\mathrm{B}}T_x)^2}\\
    & \int \mathrm{d}\epsilon_{\mathbf{k}_2} \mathrm{d}\epsilon_{\mathbf{k}_3} \mathrm{d}\theta_2 \mathrm{d}\theta_3e^{-2L^2(k'_1k_2 cos\theta_2-k'_1k_3cos\theta_3}e^{-L^2k^{'2}_1}\\
    &e^{-\beta(\epsilon_{\mathbf{k}_3}+\epsilon_{\mathrm{lp}})}\mathcal{H}(\epsilon_{\mathbf{k}_3}+\epsilon_{\mathrm{lp}}-\epsilon_{\mathbf{k}_2}).
    \end{split}
\end{equation}
The integrals are performed over the exciton energies $\epsilon_{\mathbf{k}_2,\mathbf{k}_3}$ and the angles $\theta_{2,3}$;  $\mathcal{H}$ is the Heaviside step function, and $k'^2= k^2_3-k^2_2+\frac{2m\epsilon_{\mathrm{lp}}}{\hbar^2}$.
To reduce the number of integrals we use the approximation introduced in Ref.~\cite{emil}, which assumes the electromagnetic field to be separable with a Gaussian lateral field distribution. This simplifies the expression for the light-matter coupling $G_0$, and enables to find an analytic expression for the ERC transformation matrix $|U_{0, \mathbf{k}}|^2\approx \frac{L^2}{S}e^{-L^2k^2}$, see Appendix \ref{App_B}.

\begin{figure}[h!!]
\hspace*{-0.5cm}
\includegraphics[width=0.95\linewidth]{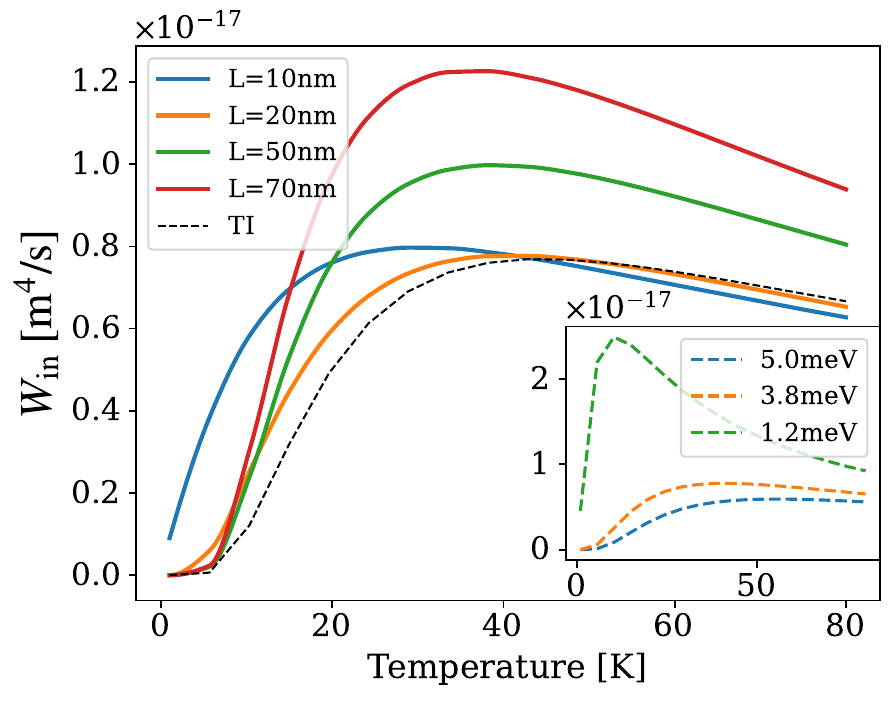}
\caption{Numerically computed in-scattering rate for $G_0 = 3.8 \mathrm{meV}$ and different optical cavity in-plane confinements. The black dotted line is the result expected for a quantum-well model where translation invariance is preserved, in which case it is possible to derive an analytic expression for $W_{\mathrm{in}}$ \cite{Porras}. The inset shows the behaviour of the scattering-in rate for $L=20\ \mathrm{nm}$, for three different coupling strengths.}
\label{fig:W_in}
\end{figure}
The example in Fig.~\ref{fig:2dmaterial} shows a so-called bowtie cavity displaying extreme dielectric confinement for which the field distribution of the quasi-normal mode can be found in Ref.~\cite{george}. We note here that the description of the transverse spatial profile of the field distribution by a Gaussian function is not necessarily a good approximation. Nevertheless, the analysis suggests that $G_0$ depends chiefly on the characteristic out-of-plane confinement length of the cavity, $L^{-1}_z$, whereas $W_0=W'_0 |X_{\mathrm{lp}}|^4$ depends mostly on the in-plane confinement length, $L$, and on none of the other length scales of the system \cite{emil}. In principle, this enables tuning the light-matter coupling strength just by changing the value of $L_z$. The three lengths scale $L$, $L_z$, $S$ that enter $W$, then play specific roles that are well distinguished. 

Eq.~\eqref{W_in_} shows the explicit dependence of the rates on the reservoir temperature. This is plotted in Fig.~\ref{fig:W_in} for different in-plane confinement lengths of the optical cavity. The scattering rates depend on the exciton occupation number $N^x_i$, which follows Maxwell-Boltzmann statistics. The dynamics of Eq.~\eqref{rate_3} then also determines the magnitude of the rates.  
The black dotted line in Fig.~\ref{fig:W_in} shows the analytic expression for $W_{\mathrm{in}}$ found in Ref.~\cite{Porras}, where the spatial confinement of the electromagnetic field is not taken into account, and it is assumed that the surface confinement is equal to $S$. Such analytic expression has the same temperature dependence obtained from our numerical simulation. Finally, the observed increase in scattering rates with varying in-plane cavity lengths can be attributed to the broadening of the Gaussian mode profile. This broadening results in an increased overlap with the Maxwell-Boltzmann distribution of the reservoir excitons. 

\begin{figure}[h!!]
\hspace*{-0.5cm}
\includegraphics[width=1\linewidth]{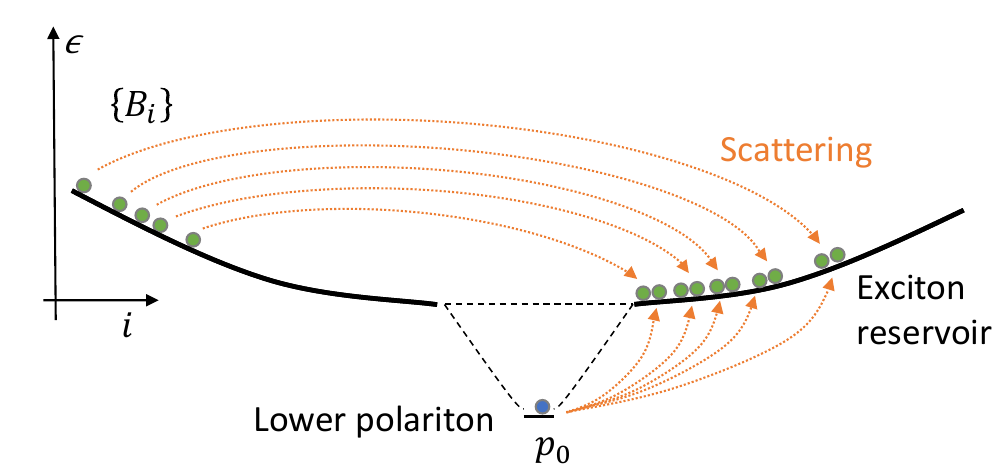}\caption{Sketch of the scattering between a lower polariton and a higher energy exciton. After the scattering, two lower exciton states are filled. The lower polaritons can scatter with any exciton in the reservoir as long as energy is conserved in the process.}
\label{fig:scattering2}
\end{figure}
As previously mentioned, the lack of conservation of in-plane momentum implies that in the expansion for $\left< F^{\dagger}_0 F_0\right>$ and $\left<F_0F^{\dagger}_0\right>$, extra terms appear that would normally cancel out for translationally invariant systems. In the following, these are labelled by $W_{\mathrm{in}_\mathrm{c}}$ and $W_{\mathrm{out}_\mathrm{c}}$; see Appendix A and B. Physically, we can understand these terms as an increase in the value of the scattering rates due to the presence of a single isolated state rather than a continuous polariton branch.
In a translationally invariant 2D system, such as a quantum well in a microcavity, Coulomb interactions conserve both energy and momentum. Two excitons in the reservoir, denoted by momentum and energy $k^x_{\mathrm{i}}$ and $\epsilon^x_i$ respectively, scatter into a lower polariton state $(k^{\mathrm{lp}}_{\mathrm{f}},\epsilon^{\mathrm{lp}}_k)$ and a higher-energy exciton $(k^x_{\mathrm{f}},\epsilon^x_{\mathrm{f}})$. Considering that the wave vector of the lower polariton is negligible compared to the scattering excitons in the reservoir, $k^{\mathrm{lp}}_{\mathrm{f}} \approx 0$, the conservation of momentum and energy results in the following conditions: 
\begin{equation}
    \begin{split}
        &2|k^x_{\mathrm{i}}|=|k^x_{\mathrm{f}}|  \rightarrow  4\epsilon^x_{\mathrm{i}}=\epsilon^x_{\mathrm{f}}\\
        & 2\epsilon^x_{\mathrm{i}}=-|\epsilon^{\mathrm{lp}}_k|+\epsilon^x_{\mathrm{f}} \rightarrow  \epsilon^x_{\mathrm{i}}=|\epsilon^{\mathrm{lp}}_k|/2, 
 \epsilon^x_{\mathrm{f}}=2|\epsilon^{\mathrm{lp}}_k| 
    \end{split}
\end{equation}
This implies that a minimum $\epsilon^x_{\mathrm{i}} = \epsilon^x_{\mathrm{lp}}/2$ is required for the scattering processes described, where now $\epsilon^x_{\mathrm{lp}}$ labels one of the states of the lower polariton branch, see Ref.~\cite{Porras}. 
These considerations do not apply to the system we study, where there is no condition on the initial energy of the scattering excitons $\epsilon^x_{\mathrm{i}}$. As illustrated in Fig.~\ref{fig:scattering2}, a polariton can interact with any exciton in the reservoir, and any exciton can scatter into the lower isolated state, provided that energy is conserved. 

\begin{figure}[h!!]
\centering'
\hspace*{0.cm}
\includegraphics[width=1\linewidth]{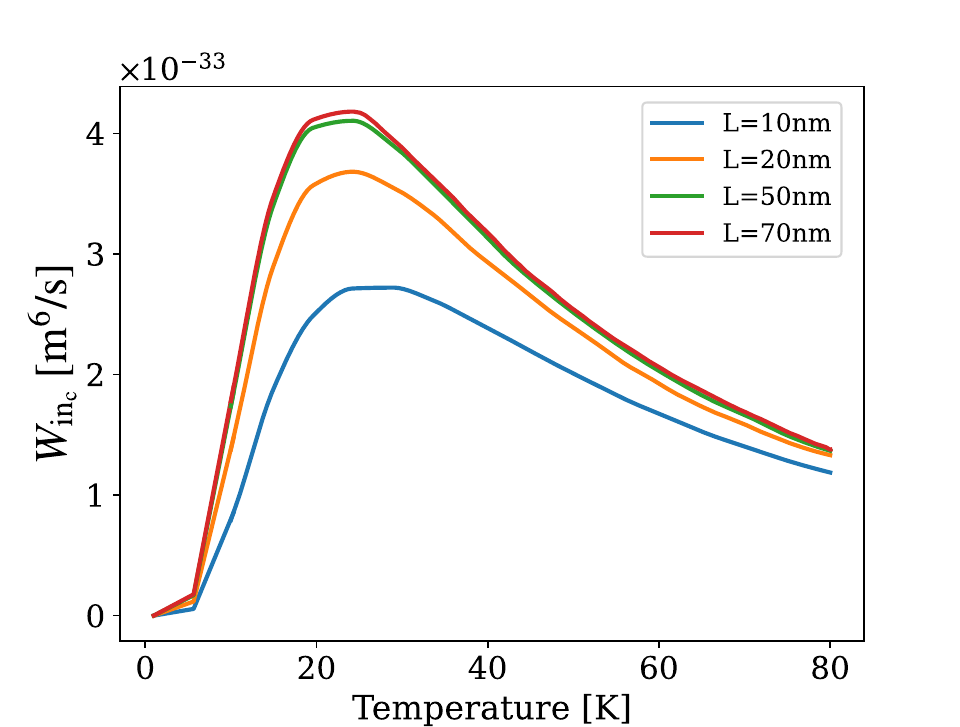}
\caption{Numerically computed $W_{\mathrm{in}_\mathrm{c}}$ scattering rate for $G_0 = 3.8\ \mathrm{meV}$ and different optical cavity in-plane confinement lengths. Increasing $L$, $W_{\mathrm{in}_\mathrm{c}}$ initially grows faster, but increasing the in-plane cavity length even more, its value remains stable.}
\label{fig:W_corr}
\end{figure}
Fig.~\ref{fig:W_corr} shows how the correction to the in-scattering rate varies with temperature for different in-plane confinement values. As the cavity size increases, the rate follows the same trend as $W_{\mathrm{in}}$. Furthermore, its value remains stable for $L =50 \ \mathrm{nm}, 70 \ \mathrm{nm}$. It follows that when $L$ becomes sufficiently large, the correction terms cease to grow and become less relevant in the system's dynamics. 

The validity of the model is subject to two important constraints. The first one pertains to the maximum allowable value of the exciton density, which is limited to less than the saturation density $n_{\mathrm{sat}} \approx 1/a^2_{\mathrm{B}} = 2\times 10^{17} \ \mathrm{m}^{-2}$. This restriction ensures that the system remains below the Mott transition and justifies the description of the excitons as interacting bosons \cite{tassone_mottdensity,Sivakumar2018-bg}. The second constraint limits the maximum allowable value of the total exciton occupation number. The population of the exciton reservoir is modelled as a Maxwell-Boltzmann distribution where quantum degeneracy effects are negligible. This requires $N^x < 1$ and from Eq.~\eqref{occupation_number} implies that $n_x < k_{\mathrm{B}} T_x \rho_x/2\pi$ \cite{Porras}. 

\subsection{\label{sec:level3}Results}
In this section, we present and discuss numerical results obtained from our model. In the simulations below, the 2D material is chosen to be Tungsten disulfide ($\mathrm{WS_2}$), for which the important material parameters are shown in Table \ref{tab:table}. For the nanocavity, the out-of-plane-confinement of the model-cavity has been adjusted to tune the value of the coupling strength $G_0$ \cite{emil}.

\begingroup
\renewcommand{\arraystretch}{1.5} 
\begin{table}[h!!]
    \centering
\begin{tabular}{||c | c | c | c | c|c||}  
 \hline  
    $a_{\mathrm{B}}[\mathrm{nm}]$  & $m_{\mathrm{h}}[m_0]$ & $m_{\mathrm{e}}[m_0]$  & $E_{\mathrm{g}}[\mathrm{eV}]$ & $E_{\mathrm{b}}[\mathrm{eV}]$ & $\tau_x[\mathrm{ps}]$\\[1.5ex] 
    \hline 
   $1.95$  & $0.34$ & $0.33$ & $2.53$ & $0.52$& $2$\\
 [1.ex]  
 \hline 
\end{tabular}
\caption{Relevant material parameters for a monolayer, $\mathrm{WS_2}$. The listed quantities are the exciton Bohr radius $a_{\mathrm{B}}$; the effective hole and electron masses $m_{\mathrm{h}}$ and $m_{\mathrm{e}}$, respectively, given in units of the free electron mass $m_0$; the band-gap energy $E_{\mathrm{g}}$, and the exciton binding energy $E_{\mathrm{b}}$ \cite{Li2014, Denning_2022}. For the exciton lifetime $\tau_x$ we refer to Ref.~\cite{Selig2016}.}
    \label{tab:table}
\end{table}
\endgroup

We study and discuss the system at steady-state as a function of the pumping and find that the evolution of the ground-state occupation exhibits a distinct kink at which it can increase by orders of magnitude with only a moderate increase in the pump rate. We regard this highly non-linear behaviour as the onset of polariton condensation. 
We pay particular attention to the role played by the temperature, differentiating the role of $T_x$ and $T_\mathrm{L}$. The dynamics of the former turn out to be strongly linked to the number of particles in the lower polariton state. By tuning the latter, we can control the onset of condensation as we vary the pumping rate. 

In steady-state, from Eqs. \eqref{rate5}, \eqref{rate_2} and \eqref{rate_3}, we get the following coupled equations
\begin{align}
\label{eq:N0ss}
    &N_0^{\mathrm{ss}}=\frac{R_{\mathrm{sp}}(n_x^{\mathrm{ss}},T_x^{\mathrm{ss}})}{\Gamma_{\mathrm{lp}}-G_{\mathrm{st}}(n_x^{\mathrm{ss}},T_x^{\mathrm{ss}})},\\
\label{eq:pxss}
    &p_x = \Gamma_{x}n_x^{\mathrm{ss}}+\frac{\Gamma_{\mathrm{lp}}N_{0}^{\mathrm{ss}}}{S},\\
\label{eq:Txss}
    &p_xk_{B}T_{\mathrm{L}}=\Gamma_{x}n_x^{\mathrm{ss}}k_BT_x^{\mathrm{ss}}+\frac{\Gamma_{\mathrm{lp}} N_0^{\mathrm{ss}}\epsilon_{\mathrm{lp}}}{S}, 
\end{align}
where $R_{\mathrm{sp}}(n_x,T_x)$ and $G_{\mathrm{st}}(n_x,T_x)$ are functions representing respectively the rate of spontaneous and stimulated scattering into the polariton branch,
\begin{align}
\label{R}
R_{\mathrm{sp}}&= n_x^2W_{\mathrm{in}}(T_x)+n_x^3W_{\mathrm{in}_\mathrm{c}}(T_x),\\
\label{G}
\begin{split}
G_{\mathrm{st}}&=n_x^2W_{\mathrm{in}}(T_x)-n_x^2W_{\mathrm{out}_\mathrm{2}}(T_x)-\\
    &-n_x^2W_{\mathrm{out}_\mathrm{c}}(T_x)-n_xW_{\mathrm{out}_\mathrm{1}}(T_x).
    \end{split}
    \end{align}
Eq. (\ref{eq:N0ss}) gives the polariton number in steady-state for a given temperature and exciton density. Eq. (\ref{eq:pxss}) represents the conservation of particles, and Eq. (\ref{eq:Txss}) corresponds to the conservation of energy. The input-output relations are found by solving these coupled nonlinear algebraic equations numerically.

\begin{figure}[h!!]
\hspace*{-.4cm}
\includegraphics[width=1.\linewidth]{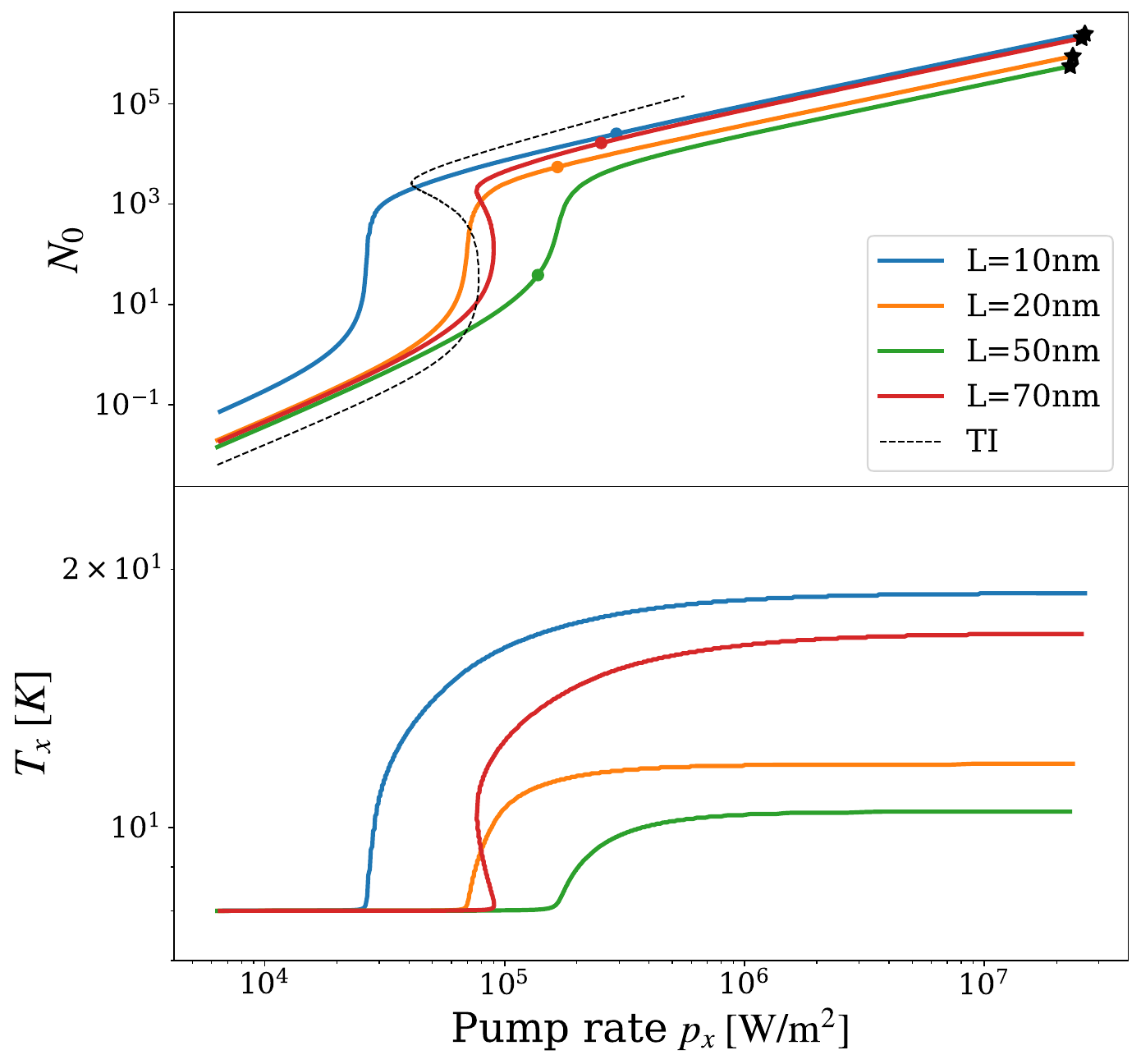}
\caption{Upper plot, polariton occupation as a function of pump power for different in-plane confinement lengths of the optical cavity, $G_0 =3.8 \ \mathrm{meV}$ and $T_{\mathrm{L}} = 8 \ \mathrm{K}$. The small dots in each curve label the point where the corresponding occupation number of the excitons $N^x \sim 1$ and Maxwell-Boltzmann distribution breaks down in this limit. The black star refers to the value corresponding to the Mott transition of $n_x$. The black dotted line represents the system's dynamics without correction terms, assuming analytic expressions for in-scattering and out-scattering rates \cite{Porras} and considering a single polariton state. The bottom plot shows the variation in the exciton reservoir temperature.}
\label{fig:N0_evolution}
\end{figure}
Fig.~\ref{fig:N0_evolution} shows the change in the steady-state values of the polariton occupation (upper plot) and the reservoir temperature (bottom plot) as a function of the input power. The four curves refer to a system with light-matter coupling $G_0 = 3.8\ \mathrm{meV}$, $T_{\mathrm{L}}=8\ \mathrm{K}$ and different in-plane lengths $L$ of the optical cavity. The extension of the lattice surface is maintained constant at $S= (15 \ \mathrm{\mu m})^2$ for the four simulations so that the same pump is required to excite the same initial density of excitons. $L$ is progressively increased, enhancing the value of the rates as shown in Fig.~\ref{fig:W_in} for $W_{\mathrm{in}}$ and Fig.~\ref{fig:W_corr} for $W_{\mathrm{in_c}}$. The upper plot in Fig.~\ref{fig:N0_evolution} shows that as $L$ increases, a higher pumping rate becomes necessary to observe a kink in the evolution of $N_0$. This continues until $L = 70 \ \mathrm{nm}$ (red curve) where the required pump decreases, bringing the curve closer to the black dotted one, which represents the system's dynamics for a single polariton state, without correction terms and assuming analytic expressions for the in-scattering rate (black dotted curve in Fig.~\ref{fig:W_in}) and the out-scattering rate \cite{Porras}. We can conclude that larger cavity sizes necessitate bigger pumping to observe a kink in $N_0$. For $L$ sufficiently large, moreover, the effect of $W_{\mathrm{in_c}}$ and $W_{\mathrm{out_c}}$ becomes progressively smaller in $N_0$ dynamics and we recover the limit where $W_{\mathrm{in}}$ and $W_{\mathrm{out_1}}$ have analytic expressions.

As seen in the bottom panel of Fig.~\ref{fig:N0_evolution}, the onset of condensation leads to an increase in the reservoir temperature because more excitons are scattered into higher energy states. This increase in the value of $T_x$ can, in turn, result in bistability in the evolution of $N_0$, as evident from the dynamics displayed by the red and black dotted curves in the top panel Fig.~\ref{fig:N0_evolution}. Further elucidation of this phenomenon will be presented in subsequent discussions.
\begin{figure}[h!!]
\hspace*{-0.5cm}
\includegraphics[width=1.1\linewidth]{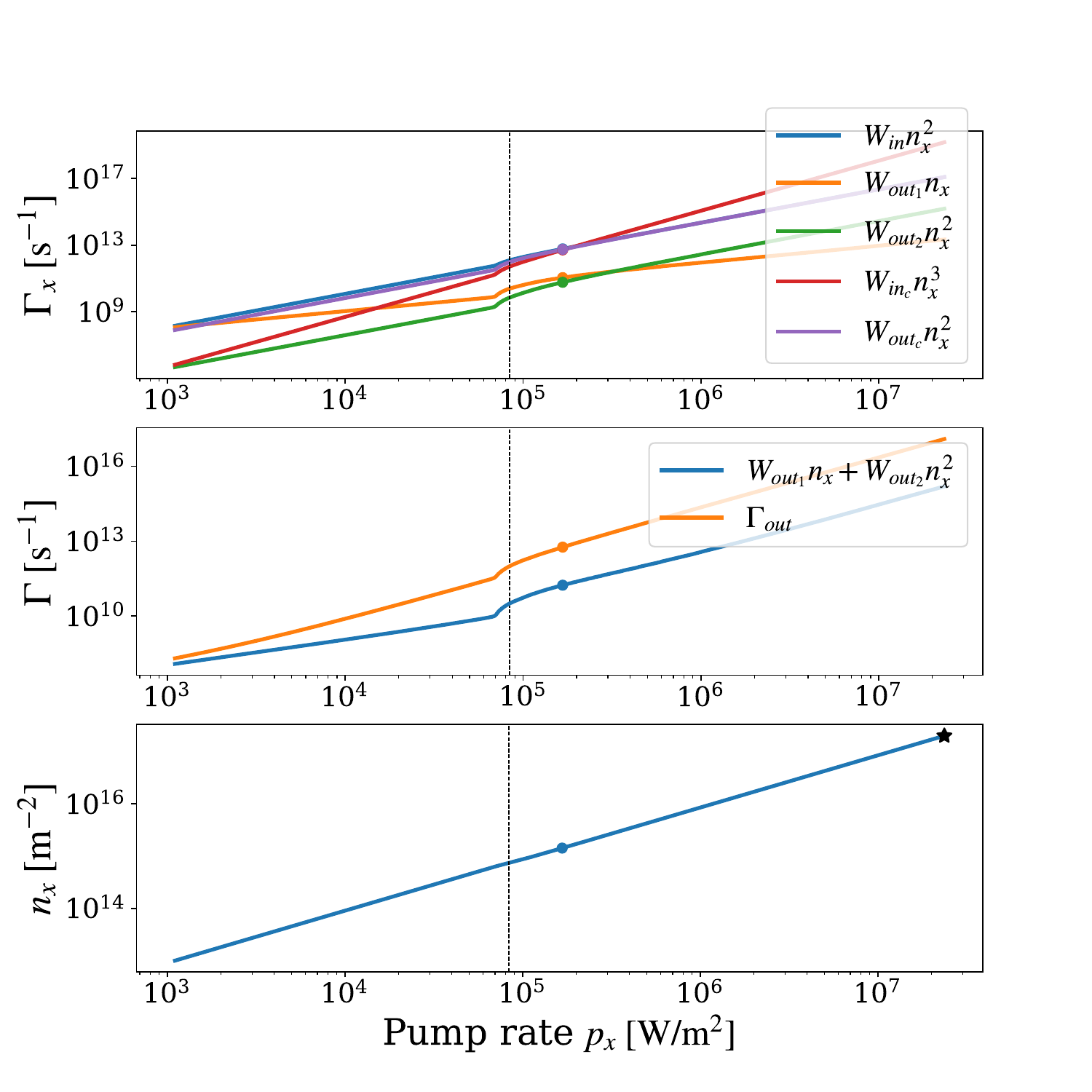}
\caption{Upper plot, the evolution of the scattering rates as a function of the pump power $p_x$ for $G_0 = 3.8\ \mathrm{meV}$ and $L=20 \ \mathrm{nm}$. Middle plot, evolution of the total scattering-out rate $\Gamma_{\mathrm{out}} (p_x)=n^2_xW_{\mathrm{out}_\mathrm{1}}+n^2_xW_{\mathrm{out}_\mathrm{2}}+ n^2_xW_{\mathrm{out_c}}$ compared to $n_xW_{\mathrm{out}_\mathrm{1}}+n^2_xW_{\mathrm{out}_\mathrm{2}}$. Bottom plot, corresponding evolution of the exciton density $n_x$. The dotted black line is the $p_x$ value corresponding to the $N_0$ threshold. The black stars indicate the onset of the Mott transition. }
\label{fig:20nm}
\end{figure}

The evolution of $T_x$ illustrates that the system behaves qualitatively differently from the traditional case of a system with translational invariance. In the latter case, the exciton temperature increases with the pumping \cite{Porras}. From Fig.~\ref{fig:N0_evolution}, however, the exciton temperature approaches a saturation limit, which depends on the coupling strength and the cavity size. This behaviour seemed rather linked to the consideration in Sec.~\ref{scattering}, where it was pointed out that there is not a minimum energy required for the particle to scatter into $p_0$. 

\begin{figure}[h!!]
\hspace*{-.5cm}
\includegraphics[width=1\linewidth]{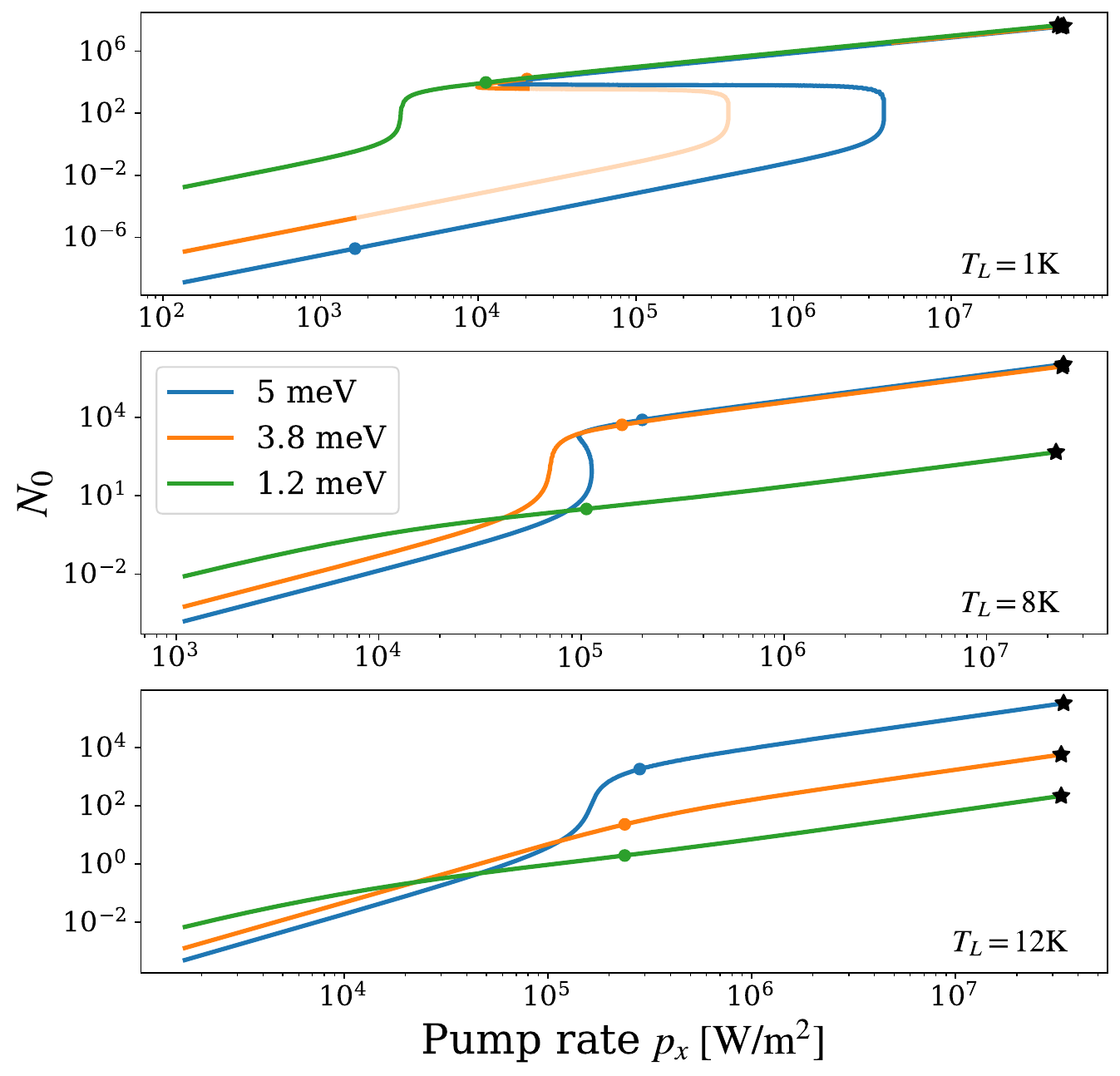}
\caption{Polariton occupation number $N_0$ as a function of the pump power $p_x$ for different coupling strengths $G_0$ and lattice temperatures. In the upper plot $T_{\mathrm{L}} = 1\ \mathrm{K}$, in the middle $T_{\mathrm{L}} = 8\ \mathrm{K}$ and in the bottom $T_{\mathrm{L}} = 12\ \mathrm{K}$. The small dots in each curve label the point where the corresponding occupation number of the excitons $N_x \sim 1$ and Maxwell-Boltzmann distribution breaks down. The black star indicates the density at which the Mott transition for the exciton reservoir occurs. Dashed regions of the curves indicate an exciton density higher than the limit imposed by Maxwell-Boltzmann when the exciton energy is still defined by the initial lattice temperature $T_{\mathrm{L}}$.}
\label{fig:comparison}
\end{figure}

The different scattering rates, together with the evolution of $n_x$ as a function of the pumping rate, are plotted in Fig.~\ref{fig:20nm}, for a cavity length of $L=20\ \mathrm{nm}$. The black vertical line shows the pumping rate needed to reach the threshold. Here, we define the threshold as the condition for which, for a given temperature, the gain equals the losses, see Eq.~\eqref{threshold_n} below. From the top plot of Fig.~\ref{fig:20nm}, we observe that $W_{\mathrm{in_c}}n^3_x$ grows progressively faster than the other rates as a function of the pump. The total Coulomb out-scattering of the polariton state is also modified by $W_{\mathrm{out}_\mathrm{c}}$, as seen in the centre plot. In Sec.~\ref{sec:level1}, we interpret the emergence of the correction terms as an enhancement of the overall scattering rates originating from a single polariton state; see Fig.~\ref{fig:scattering2}. To further illustrate this behaviour, Fig.~\ref{fig:20nm} shows the evolution of $\Gamma_{\mathrm{out}}(p_x)=n_xW_{\mathrm{out}_\mathrm{1}}+n^2_xW_{\mathrm{out}_\mathrm{2}}+ n^2_xW_{\mathrm{out_c}}$ compared to $n_xW_{\mathrm{out}_\mathrm{1}}+n^2_xW_{\mathrm{out}_\mathrm{2}}$. From these comparisons, it becomes evident that the correction rates, in this case $n_x^3W_{\mathrm{out_c}}$, become increasingly more important as the system approaches the threshold, emphasizing the necessity of accounting for the loss of translation invariance to describe the system accurately. We therefore conclude that we cannot neglect these terms since they contribute in a non-negligible way to the dynamics.

To gain deeper insights into the dynamics, we compare the steady-state values of the lower polariton population $N_0$ for three different coupling strengths while keeping the confinement length fixed at $20 \ \mathrm{nm}$, see Fig.~\ref{fig:comparison}. This analysis emphasizes the crucial role played by the lattice temperature in defining a kink and the presence of hysteresis. In the upper plot of the figure, in which we consider the case in which the initial temperature is set to $T_{\mathrm{L}} = 1\ \mathrm{K}$, a distinct kink is observed. As we progressively increase $T_{\mathrm{L}}$, however, the middle and bottom plots show that only the systems with higher values of $G_0$ exhibit a deviation from linear evolution.
In the present model, as well as in a translation-invariant case, a higher coupling strength, meaning an increase in the value of $|\epsilon_{\mathrm{lp}}|$, leads to the scattering of reservoir excitons to higher-energy states, raising the reservoir temperature even more. Based on this understanding, we can make some conclusions on the saturation temperature earlier introduced in Fig.~\ref{fig:N0_evolution}. This value is expected to grow with $G_0$ as well. Thus, we expect to observe the same dynamics at higher temperatures as $G_0$ increases. 

\begin{figure}[h!!]

\includegraphics[width=1.\linewidth]{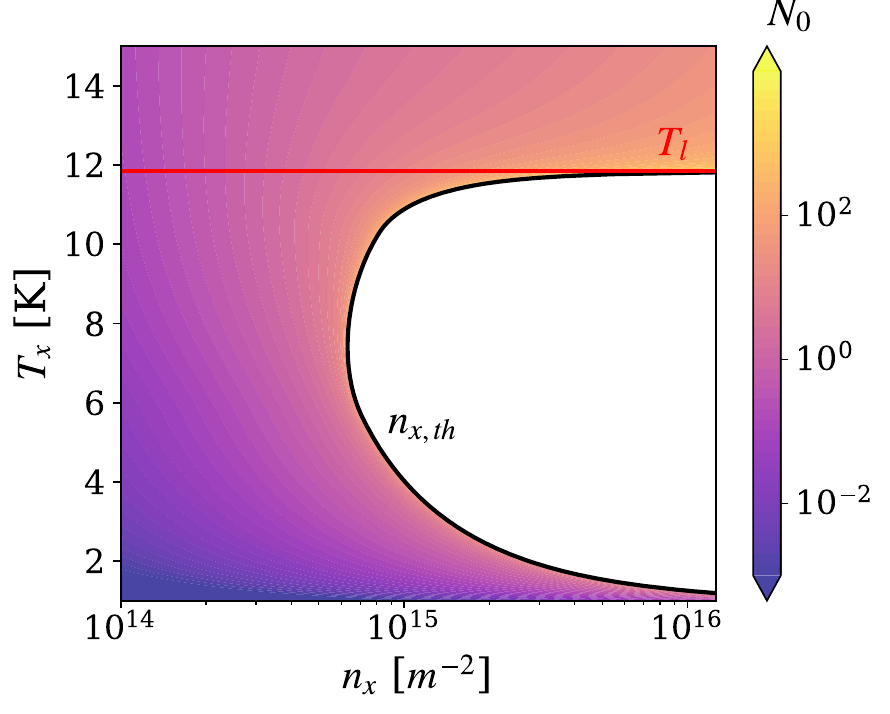}
\caption{Colour map of the steady-state value of the ground state population, $N_0^{\mathrm{ss}}$ versus the exciton reservoir density $n_x$ and temperature $T_x$. The black line shows the threshold exciton density $n_{x,\mathrm{th}}(T_x)$, where $N_0^{\mathrm{ss}}$  diverges. The red line,  $T_{\mathrm{l}} \simeq 12\ \mathrm{K}$, defines the limit temperature, behind which the system cannot reach a threshold anymore.}
\label{fig:threshold}
\end{figure}
A formal definition for the limit temperature $T_{\mathrm{l}}$ can be obtained by studying the polariton distribution at steady-state. Equation \eqref{rate5} can be rewritten as
\begin{equation}
     \frac{dN_0}{dt}=R-\Gamma_{\mathrm{lp}}N_0,
\end{equation}
where the polariton generation rate $R$ containes a spontaneous contribution and a stimulated contribution
\begin{equation}
    R = R_{\mathrm{sp}}+G_{\mathrm{st}}N_0.
\end{equation}
both depending on $T_x$ and $n_x$ and previously defined in Eqs. \eqref{R} and \eqref{G}.
In steady-state, if $n_x$ and $T_x$ are known, the polariton population is given by 
\begin{equation} \label{N0ss}N_0^{\mathrm{ss}}=\frac{R_{\mathrm{sp}}}{\Gamma_{\mathrm{lp}}-G_{\mathrm{st}}},
\end{equation}
which is plotted in Fig.~\ref{fig:threshold} as a function of $n_x$ and $T_x$. For small pumping, we can approximate
\begin{equation}
    N_0^{\mathrm{ss}}\approx R_{\mathrm{sp}}/\Gamma_{\mathrm{lp}}.
\end{equation}
This approximation breaks down near the threshold. For a given temperature $T_x$, we define the threshold exciton density $n_{x,\mathrm{th}}$ as the point where the gain equals the losses
\begin{equation}\label{threshold_n}
 G_{\mathrm{st}}(n_{x,\mathrm{th}}(T_x),T_x)-\Gamma_{\mathrm{lp}}=0,
\end{equation}
which is plotted as the black line in Fig.~\ref{fig:threshold}.
Solving Eq.~\eqref{threshold_n} for $n_{x,\mathrm{th}}$, we get the solution
\begin{equation}\footnotesize
    n_{x,\mathrm{th}}=\frac{W_{\mathrm{out}_\mathrm{1}}}{2W_{2}}+\sqrt{\left(\frac{W_{\mathrm{out}_\mathrm{1}}}{2W_2}\right)^2+\frac{\Gamma_{\mathrm{lp}}}{W_2}},
\end{equation}
with the requirement for a threshold, i.e. a positive and real solution, being $W_2 = W_{\mathrm{in}}-W_{\mathrm{out}_\mathrm{2}}-W_{\mathrm{out}_\mathrm{c}}>0$.
At $T_x=T_{\mathrm{l}}$ the threshold diverges, coinciding with the point where $W_2\rightarrow 0$, which is shown as the red line in Fig.~\ref{fig:threshold}. Above this temperature, $G_{\mathrm{st}}$ is negative for all $n_x$ and we do not obtain stimulated emission into the polariton branch. Thus, $T_{\mathrm{l}}$ is defined by
\begin{equation}
    W_2(T_{\mathrm{l}})\equiv 0.
\end{equation}
\begin{figure}[h]
\hspace*{-1.7cm}
\includegraphics[width=1.4\linewidth]{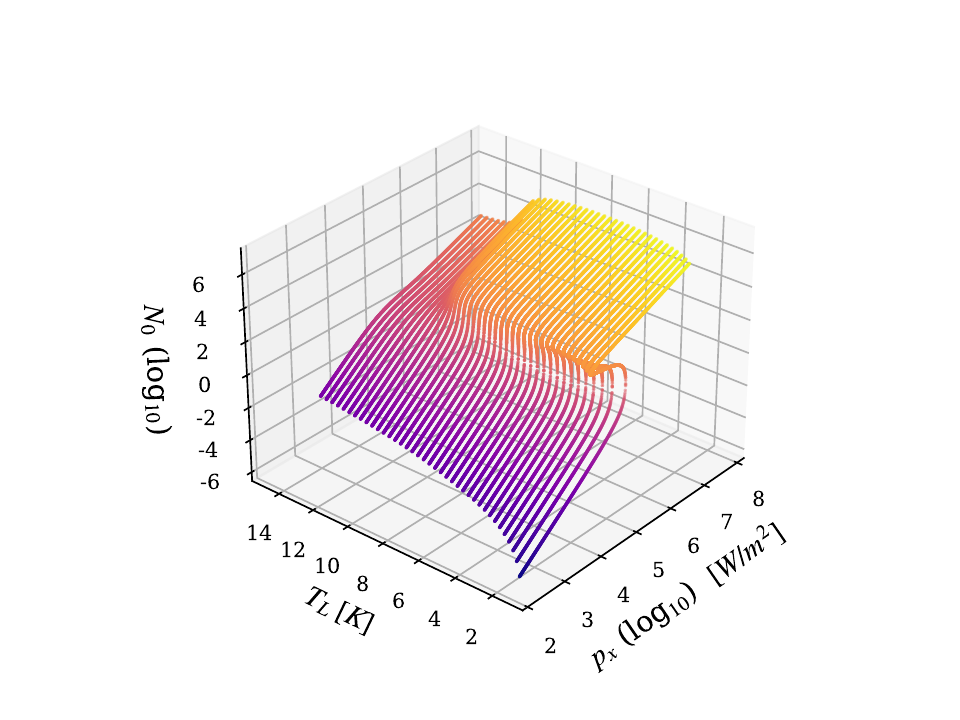} \caption{Three-dimensional plot of the steady-state values of $N_0$ versus $p_x$ and $T_\mathrm{L}$, for fixed $G_0 =3.8\ \mathrm{meV}$ and $L=20 \ \mathrm{nm}$. The colour map represents the variation in the polariton occupation number, lighter colours correspond to higher values of $N_0$. There are three phases corresponding to bistability, threshold and not-threshold. These are reached by increasing the value of the lattice temperature $T_{\mathrm{L}}$ and letting $N_0$ evolve as a function of the pumping $p_x$.}
\label{fig:phasediagram}
\end{figure}

This result can be recognized in the three-dimensional plot in Fig.~\ref{fig:phasediagram}, which shows the steady-state values of $N_0$ versus $p_x$ and $T_\mathrm{L}$, for fixed $G_0 =3.8\ \mathrm{meV}$ and $L=20 \ \mathrm{nm}$. For smaller values of $T_{\mathrm{L}}$, the dynamics of $N_0$ show initial bistability followed only by a kink. Increasing the temperature above $T_{\mathrm{L}} \simeq 12 \ \mathrm{K}$, even the kink in $N_0$ disappears, and its growth as a function of the pump becomes linear in a log-log scale. We identify $T_{\mathrm{l}}$ as the value of the temperature marking the boundary between these two regions.

\subsubsection{Discussion} 
The simulations show that a kink in the polariton population is not always observed, even when the occupation number reaches values larger than one, and we expect final state stimulation as a consequence of the bosonic nature of the quasi-particles \cite{Huang2000}. The stimulated scattering into the ground state depends on the initial lattice temperature, and above some limit value $T_{\mathrm{l}}$, the dynamics do not show any threshold, as is visible in Fig.~\ref{fig:phasediagram}. We also point out that for a sufficiently low initial temperature, the exciton reservoir evolves to $T_{\mathrm{l}}$ and maintains this temperature regardless of how much the pumping rate is increased. As noted above, this is a consequence of the loss of translation invariance in the system, leading to a result different from the one predicted for quantum wells \cite{Porras}.

When the initial lattice temperature is further decreased, Fig.~\ref{fig:phasediagram} shows the appearance of hysteresis. Bistability is often observed in polariton systems that are resonantly pumped \cite{Baas2004-iu}. This originates from a nonlinear frequency shift of the polariton mode along with near-resonant, but slightly detuned, pumping. Initially, the threshold is high because of the detuning, but if a polariton population is built up, the polariton energy shifts into resonance with the pump, decreasing the threshold \cite{Sanvitto2016}. We consider a non-resonant pump. Also, the magnitude of $G_0$, which defines $|\epsilon_{\mathrm{lp}}|$, is fixed in the beginning by $L_z$ \cite{emil}, so the energy of the lower polariton state is not shifting. Instead, the presence of bistability appears to be related to the dynamics of $T_x$. In particular, the dependence of the scattering rates on the exciton temperature induces a temperature-dependent threshold $n_{x,\mathrm{th}}(T_x)$. Bistability occurs if $n_{x,\mathrm{th}}(T_x)$ \textit{decreases} as $T_x$ increases above the initial lattice temperature. This becomes more clear from Fig.  \ref{fig:threshold} where bistability can be observed if $T_{\mathrm{L}}$ is chosen such that there exists a region of $T_x>T_{\mathrm{L}}$  where $n_{x,\mathrm{th}}(T_x)<n_{x,\mathrm{th}}(T_{\mathrm{L}})$. For the parameters in Fig.~\ref{fig:threshold}, bistability occurs for $T_{\mathrm{L}}\lesssim 7.5$ K.

From these considerations, it will be relevant to derive the evolution of the emission linewidth from the first-order two-time correlation function to understand the system better and eventually predict the observation of a lasing-like emission spectrum. Such calculations, however, are outside the scope of the present article and are left for a later study.

\subsection{\label{sec:level5}Conclusions} 
We have developed a model to describe the interaction between light and matter in a 2D material system coupled with a nanoscale electromagnetic resonator. Light-matter interaction breaks the translational symmetry of excitons in the 2D sheet, leading to the emergence of a localized polariton state. Our study reveals that the formation of a macroscopic population in the localized polariton state is influenced by correction terms addressing the breakdown of translational symmetry. As the spatial width of the cavity mode increases, Coulomb scattering rates rise, while the correction terms saturate, progressively exerting less influence on the system's dynamics. Numerical simulations demonstrate the emergence of a threshold, bistability, and hysteresis. These phenomena can be tuned by the initial lattice temperature while keeping the in-plane cavity size and light-matter interaction strength constant. It is noteworthy that we observe hysteresis under non-resonant pumping, which is distinct from the hysteresis observed in larger polariton microcavities under resonant pumping. Thus, the cause for bistability in this case is attributed to different factors.

\subsection{ACKNOWLEDGMENTS}
The authors thank George Kountouris for providing Fig.~\ref{fig:2dmaterial}. This paper was supported by the Danish National Research Foundation through NanoPhoton-Centerthe  for Nanophotonics, Grant No. DNRF147. \newline
E. V. D. acknowledges support from Independent Research Fund Denmark through an International Postdoc Fellowship (Grant No. 0164-00014B)

\begin{appendix}
\section{Expectation values}\label{App_A}
The scattering rates are computed from Wick's theorem by expanding the six-operator expectation value in Eq.\eqref{rate_derivation}. For this computation, we make the following remarks: a) the 2D sheet is sufficiently big such that each exciton reaction coordinate coefficient can be mapped into a momentum $\mathbf{k}$. It follows that $U_{i,\alpha,\mathbf{k}}= \delta_{i,\mathbf{k}}$. b) The Coulomb interaction matrix is assumed to be real $W^*_{\mathbf{k},\mathbf{k}',\mathbf{q}}=W_{\mathbf{k},\mathbf{k}',\mathbf{q}}$. c) The approximation $W_{\mathbf{k},\mathbf{k}',\mathbf{q}}\simeq W_{000}$ still holds. d) The matrices for the change of basis in the ERC are unitary, giving a delta function that allows us to consider only some of the terms in the long expansion of Wick's theorem. e) Effects of intervalley Coulomb exchange are neglected \cite{emil}.
We can then expand
\begin{equation}
    \begin{split}\footnotesize
    &\left<F^{\dagger}_0F'_0\right>= \sum_{ijhi'j'h'}\mathcal{U}_{i,j,h}\mathcal{U}^*_{i',j',h'}\left<B^{\dagger}_iB_jB_hB'^{\dagger}_{h'}B'^{\dagger}_{j'}B'_{i'}  \rho_{\mathrm{R}}\right>=\\
        &=\hbar^2\sum_{\substack{\mathbf{k}_1,\mathbf{k}_2,\mathbf{k}_3 \\ \mathbf{k}_4,\mathbf{q},\mathbf{q}' \\ ijhi'j'h'}}e^{\frac{i(t-s)}{\hbar}(\epsilon_{h'}+\epsilon_{j'} -\epsilon_{i'})}W^*_{\mathbf{k}_1,\mathbf{k}_2,\mathbf{q}}W_{\mathbf{k}_3,\mathbf{k}_4,\mathbf{q}'}\\
        &U_{0,\mathbf{k}_1+\mathbf{q}}U_{i,\mathbf{k}_2-\mathbf{q}}U^*_{j,\mathbf{k}_1}U^*_{h,\mathbf{k}_2}U_{h',\mathbf{k}_3+\mathbf{q}}U_{j',\mathbf{k}_4-\mathbf{q}}U^*_{i',\mathbf{k}_3}U^*_{0,\mathbf{k}_4}\\
        & \big[ \delta_{ij}N_i\big[(1+N_h)N_j'\delta_{hh'}\delta_{j'i'}+(1+N_h)N_h'\delta_{hj'}\delta_{h'i'}\big]+\\
        &+\delta_{ih}N_i \big[(1+N_j)N_{j'}\delta_{jh'}\delta_{j'i'}+(1+N_j)N_{h'}\delta_{jj'}\delta_{h'i'}\big]+\\
        &+\delta_{ii'}N_i\big[(1+N_j)(1+N_h)\delta_{jh'}\delta_{j'h}+\\
        &+(1+N_j)(1+N_h)\delta_{jj'}\delta_{hh'}\big] \big]
    \end{split}
\end{equation}
The same expansion is performed for the term, $\left<F_0F^{'\dagger}_0\right>$.
It follows that the expression for the rate of injection into the lower polariton state is 
\begin{equation}
\begin{split}\footnotesize
     &\left<F_0F^{\dagger}_0\right>\Big(e^{i\frac{(t-s)\epsilon_{\mathrm{lp}}}{\hbar}}+e^{-i\frac{(t-s)\epsilon_{\mathrm{lp}}}{\hbar}}\Big)=\\
        &(e^{i\frac{(t-s)\epsilon_{\mathrm{lp}}}{\hbar}}+e^{-i\frac{(t-s)\epsilon_{\mathrm{lp}}}{\hbar}})\hbar^2\sum_{\mathbf{k}_1,\mathbf{k}_2,\mathbf{k}_3}\big[\big[N_{\mathbf{k}_1}N_{\mathbf{k}_2}N_{\mathbf{k}_3}|U_{0,\mathbf{k}_1}|^2\\
        &(W_{\mathbf{k}_1,\mathbf{k}_2,\mathbf{0}}W_{\mathbf{k}_3,\mathbf{k}_1,\mathbf{0}}+W_{\mathbf{k}_1,\mathbf{k}_2,\mathbf{0}}W_{\mathbf{k}_3,\mathbf{k}_1,\mathbf{k}_1-\mathbf{k}_3})+\\
        &+\big[N_{\mathbf{k}_1}N_{\mathbf{k}_2}N_{\mathbf{k}_3}|U_{0,\mathbf{k}_2}|^2(W_{\mathbf{k}_1,\mathbf{k}_2,\mathbf{k}_2-\mathbf{k}_1}W_{\mathbf{k}_3,\mathbf{k}_2,\mathbf{0}}+\\&+W_{\mathbf{k}_1,\mathbf{k}_2,\mathbf{k}_2-\mathbf{k}_1}W_{\mathbf{k}_3,\mathbf{k}_2,\mathbf{k}_2-\mathbf{k}_3})\big]+
     \big[N_{\mathbf{k}_1}N_{\mathbf{k}_2}(1+N_{\mathbf{k}_3})\\
        &|U_{0,\mathbf{k}_1+\mathbf{k}_2-\mathbf{k}_3}|^2(W_{\mathbf{k}_1,\mathbf{k}_2,\mathbf{k}_2-\mathbf{k}_3}W_{\mathbf{k}_3,\mathbf{k}_1+\mathbf{k}_2-\mathbf{k}_3,\mathbf{k}_1-\mathbf{k}_3}\\&+W_{\mathbf{k}_1,\mathbf{k}_2,\mathbf{k}_2-\mathbf{k}_3}W_{\mathbf{k}_3,\mathbf{k}_1+\mathbf{k}_2-\mathbf{k}_3,\mathbf{k}_2-\mathbf{k}_3})\big]\big]
\end{split}
\end{equation}
A similar expression is recovered for the out-scattering rate from $\left<F^{\dagger}_0F_0\right>\Big(e^{i\frac{(t-s)\epsilon_{\mathrm{lp}}}{\hbar}}+e^{-i\frac{(t-s)\epsilon_{\mathrm{lp}}}{\hbar}}\Big)$.

\section{Scattering rates}\label{App_B}
We follow the steps performed in Ref.~\cite{emil} to simplify the expression for $U_{0\mathbf{k}}$ of the 2D sheet. The coupling term can be expanded into
\begin{equation}\label{coupling} \begin{split}
    G_0 = &\sqrt{\sum_{\mathbf{k}}{|g_{\mathbf{k}}|^2}} \rightarrow \\ &\hbar g_{\mathbf{k}}=-\frac{e_0}{m_0}\sqrt{\frac{\hbar}{2\epsilon_0\omega_{\mathrm{c}}}}\mel{\Phi_{\mathbf{k}}}{\sum_i \tilde{\mathbf{F}}_{\mathrm{c}}(\mathbf{r}_i)\hat{\mathbf{p}}_i}{0}.
    \end{split}
\end{equation}
$g_{\mathbf{k}}$ is the interaction strength between an exiton and the resonant electromagnetic field evaluated in the fermionic space. $e_0$ and $m_0$ are the free electron charge and mass, $\epsilon_0$ is the electric permittivity of free space, $\omega_{\mathrm{c}}$ is the frequency of the cavity mode, $\Phi_{\mathbf{k}}$ is the Wannier-Mott exciton states defined as a momentum superposition of electron-hole pairs. The mode profile for the electric field mode of the optical cavity is $\tilde{\mathbf{F}}_{\mathrm{c}}(\mathbf{r}_i,t)=\tilde{\mathbf{F}}_{\mathrm{c}}(\mathbf{r}_i)\delta(t)$, which defines local and non retarded coupling dynamics, $\hat{\mathbf{p}}_i$ is the particle momentum. Solving the bracket in Eq.~\eqref{coupling} through the Slater-Condon rules \cite{slater} and making some further manipulations \cite{emil}, we get to the expression for the coupling strength
\begin{equation}\label{simple}
\hbar g_{\mathbf{k}}=-\frac{e_0}{m_0}\sqrt{\frac{\hbar}{\pi\epsilon_0\omega_{\mathrm{c}}a^2_{\mathrm{B}}S}}\int d^2\mathbf{r}e^{-i\mathbf{k}\mathbf{r}}\tilde{\mathbf{F}}_{\mathrm{c}}(\mathbf{r},z_0)\mathbf{p}_{\mathrm{cv}}.
\end{equation}
$S$ is the surface area of the 2D sheet, $a_{\mathrm{B}}$ is the Bohr radius. The 2D sheet is located at $z = z_0$, and the integral is performed over the infinite extent of the 2D material. Finally, $\mathbf{p}_{\mathrm{cv}}= \int_{V_{\mathrm{uc}}}d^3\mathbf{r}u^*_{\mathrm{c}}\hat{\mathbf{p}}u_{\mathrm{v}}(\mathbf{r})$ is the matrix element of the exciton mode, with $u_i(\mathbf{r})$ a Bloch function, the subscript $i = \mathrm{c}, \mathrm{v}$ the conduction and valence band and $V_{\mathrm{uc}}$ the volume of the unit cell. From the exciton reaction coordinates 
\begin{equation}\label{spectral}
    G_0 = \sqrt{\sum_{\mathbf{k}}{|g_{\mathbf{k}}|^2}}= \sqrt{\int J(\mathbf{\omega})d\mathbf{\omega}},
\end{equation}
where $J(\mathbf{\omega})$ is the spectral density of the exciton states of the residual environment coupled to the exciton reaction coordinate $B_0$ \cite{emil}. When combining Eq.~\eqref{simple} with Eq.~\eqref{spectral}, the coupling strength becomes
   \begin{equation}
    \hbar G_0 = \sqrt{\frac{\hbar e^2_0}{\pi\epsilon_{0}m^2_0\omega_{\mathrm{c}} a^2_{\mathrm{B}}}\int d^2\mathbf{r}|\tilde{F}_{\mathrm{c}}(\mathbf{r},z_0)\mathbf{p}_{\mathrm{cv}}|^2}.
\end{equation}

This expression can be widely simplified for a localized and separable mode profile
\begin{equation}
    \hbar G_0 = \sqrt{\frac{\hbar e^2_0|\mathbf{n}\mathbf{p}_{\mathrm{cv}}|^2}{\pi\epsilon_{0} m^2_0\omega_{\mathrm{c}} a^2_{\mathrm{B}}L_z}}.
\end{equation} 
$\mathbf{n}$ is the unit polarization vector of the mode, and the out-of-plane confinement length 
\begin{equation}
    L_z= \frac{\epsilon_{\mathrm{eff}}\int d_z|\tilde{F}(z)|^2}{|\tilde{F}_z(z_0)|^2}
\end{equation}
obeying the normalization requirement $\epsilon_{\mathrm{eff}}\int d_z|\tilde{F}(z)|^2=1$, with $\epsilon_{\mathrm{eff}}$ an effective dielectric constant to account for the dielectric response of the surrounding material. 

Also, if the later field distribution defined by the electromagnetic resonator is assumed to be Gaussian  with a confinement length scale $L$ 
\begin{equation}
    F_{//}(x, y)= \frac{e^{-(x^2+y^2)/(2L^2)}}{L\sqrt{\pi}},
\end{equation}
the exciton-mode coupling term reads
\begin{equation}
    g_{\mathbf{k}}=-\sqrt{\frac{4e^2_0|\mathbf{n}\mathbf{p}_{\mathrm{cv}}|^2L^2}{\hbar \epsilon_{0} m^2_0\omega_{\mathrm{c}} a^2_{\mathrm{B}}S}}\tilde{F}_z(z_0)e^{-\frac{1}{2}kL}.
\end{equation}
It follows that the expression for $|U_{0, \mathbf{k}}|^2\sim \frac{L^2}{S}e^{-L^2k^2}$. 

\begin{figure}[]
\hspace*{-0.5cm}
\includegraphics[width=0.85\linewidth]{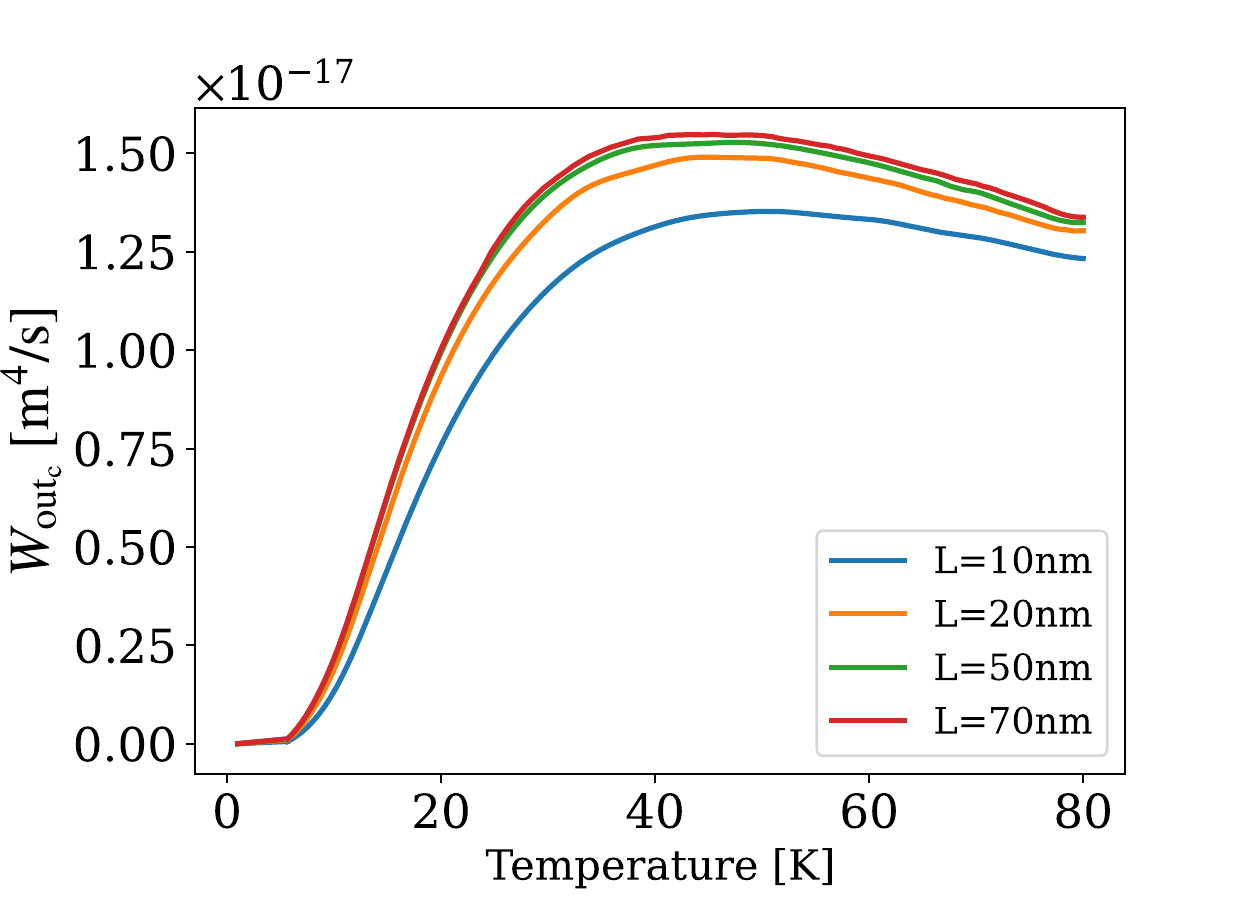}
\caption{Numerically computed out-scattering rate, $W_{\mathrm{out}_\mathrm{c}}$, for $G_0 = 3.8\ \mathrm{meV}$ and different optical cavity in-plane confinements.}
\label{fig:W_out_correction}
\end{figure}

From these results, converting the sum over $\mathbf{k}$ in Eq.~\eqref{rate_2} into an integral $\sum_{\mathbf{k}} \rightarrow \frac{S}{(2\pi)^2}\int d^2\mathbf{k}$ and solving, the following expression for the rate equations are recovered
 \begin{equation}
    \begin{split}
    W_{\mathrm{in}}=&\frac{4\hbar |W_{000}|^2|X_{\mathrm{lp}}|^2L^2 S^2 \rho_x}{2\pi (k_{\mathrm{B}}T_x)^2} \int \mathrm{d}\epsilon_{\mathbf{k}_2} \mathrm{d}\epsilon_{\mathbf{k}_3} \mathrm{d}\theta_2 \mathrm{d}\theta_3\\
    &e^{-2L^2(k'_1k_2 cos\theta_2-k'_1k_3cos\theta_3}e^{-L^2k^{'2}_1}e^{-\beta(\epsilon_{\mathbf{k}_3}+\epsilon_{\mathrm{lp}})}\\
    &\mathcal{H}(\epsilon_{\mathbf{k}_3}+\epsilon_{\mathrm{lp}}-\epsilon_{\mathbf{k}_2}),
    \end{split}
\end{equation}
\begin{figure}[h!!]
\hspace*{-0.6cm}
\includegraphics[width=0.76\linewidth]{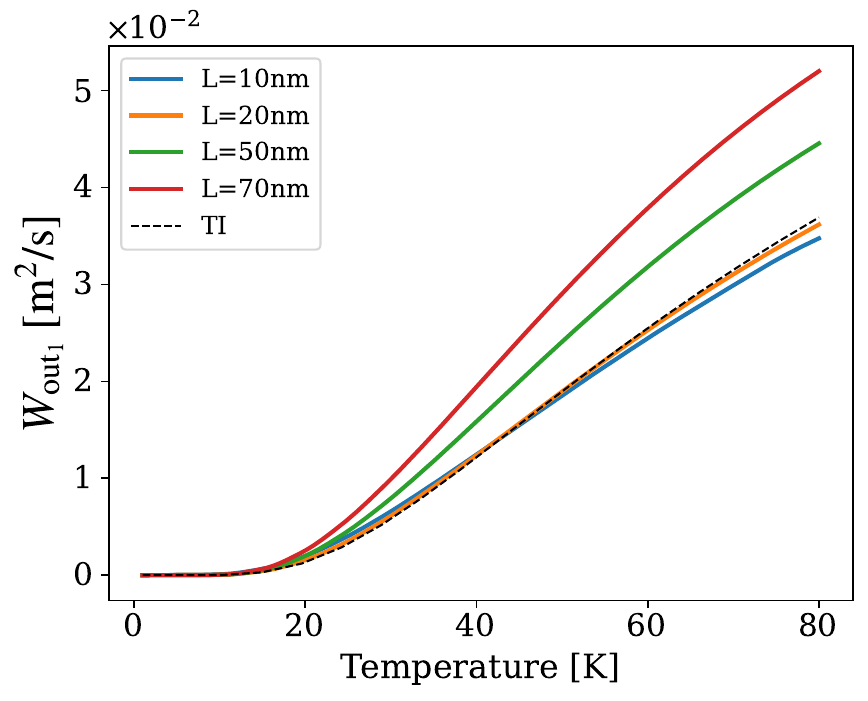}
\caption{Numerically computed out-scattering rate, $W_{\mathrm{out}_\mathrm{1}}$, for $G_0 = 3.8\ \mathrm{meV}$ and different optical cavity in-plane confinement lengths. The black dotted line is the result expected for a quantum-well model with translational invariance, in which case it is possible to derive an analytic expression for $W_{\mathrm{out}_\mathrm{1}}$ \cite{Porras}.}
\label{fig:W_out1}
\end{figure}
\begin{figure}

\includegraphics[width=0.85\linewidth]{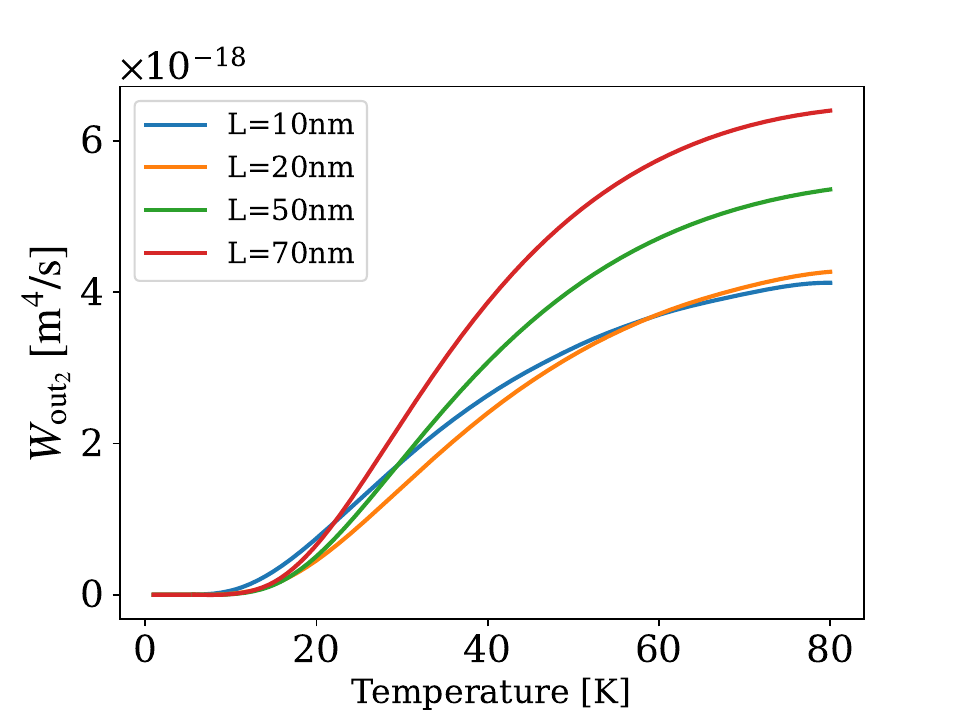}
\caption{Numerically computed out-scattering rate, $W_{\mathrm{out}_\mathrm{2}}$, for $G_0 = 3.8\ \mathrm{meV}$ and different optical cavity in-plane confinements.}
\label{fig:W_out_2}
\end{figure}
\begin{equation}
    \begin{split}
    W_{\mathrm{out}_\mathrm{1}}&=\frac{4\hbar |W_{000}|^2|X_{\mathrm{lp}}|^2L^2 S^2 \rho^2_x}{(2\pi)^2 k_{\mathrm{B}}T_x} \int  \mathrm{d}\epsilon_{\mathbf{k}_2} \mathrm{d}\epsilon_{\mathbf{k}_3} \mathrm{d}\theta_2 \mathrm{d}\theta_3\\
    & e^{-L^2|\mathbf{k'}_1+\mathbf{k}_2+\mathbf{k}_3|^2} e^{-\beta\epsilon_{\mathbf{k}_3}}\mathcal{H}(\epsilon_{\mathbf{k}_3}+\epsilon_{\mathrm{lp}}-\epsilon_{\mathbf{k}_2}).
\end{split}
\end{equation}
The variation of $W_{\mathrm{out}_\mathrm{1}}(T)$ with the temperature is plotted in Fig.~\ref{fig:W_out1}.
\begin{equation}
    \begin{split}
    W_{\mathrm{out}_\mathrm{2}}&=\frac{4\hbar |W_{000}|^2|X_{\mathrm{lp}}|^2L^2 S^2 \rho_x}{2\pi (k_{\mathrm{B}}T_x)^2} \int \mathrm{d}\epsilon_{\mathbf{k}_2}\mathrm{d}\epsilon_{\mathbf{k}_3} \mathrm{d}\theta_2 \mathrm{d}\theta_3\\
    &e^{-L^2|\mathbf{k'}_1+\mathbf{k}_2+\mathbf{k}_3|^2}\big( e^{-\beta(\epsilon_{\mathrm{lp}}+2\epsilon_{\mathbf{k}_3}-\epsilon_{\mathbf{k}_2})}+e^{-\beta(\epsilon_{\mathbf{k}_3}+\epsilon_{\mathbf{k}_2})}\big)\\
    &\mathcal{H}(\epsilon_{\mathbf{k}_3}+\epsilon_{\mathrm{lp}}-\epsilon_{\mathbf{k}_2}),
\end{split}
\end{equation}
whose evolution as a function of the temperature is plotted in Fig.~\ref{fig:W_out_2}.
\begin{equation}
    \begin{split}
    W_{\mathrm{in_c}}&=\frac{8\hbar |W_{000}|^2|X_{\mathrm{lp}}|^2L^2 S^2 (2\pi)^2}{(k_{\mathrm{B}}T_x)^3} \int \mathrm{d}\epsilon_{\mathbf{k}_2} \mathrm{d}\epsilon_{\mathbf{k}_3} e^{-L^2k'^{2}_1} \\&e^{-\beta(2\epsilon_{\mathbf{k}_3}+\epsilon_{\mathrm{lp}})}
    \mathcal{H}(\epsilon_{\mathbf{k}_3}+\epsilon_{\mathrm{lp}}-\epsilon_{\mathbf{k}_2}),\\
    \end{split}
\end{equation}
\begin{equation}
    \begin{split}
    W_{\mathrm{out_c}}&=\frac{8\hbar |W_{000}|^2|X_{\mathrm{lp}}|^2L^2 S^2 2\pi \rho_x }{(k_{\mathrm{B}}T_x)^2} \int \mathrm{d}\epsilon_{\mathbf{k}_2} \mathrm{d}\epsilon_{\mathbf{k}_3} e^{-L^2k^{'2}_1} \\
    &e^{-\beta(\epsilon_{\mathbf{k}_2}+\epsilon_{\mathbf{k}_3})}\mathcal{H}(\epsilon_{\mathbf{k}_3}+\epsilon_{\mathrm{lp}}-\epsilon_{\mathbf{k}_2}),
    \end{split}
\end{equation}
with
\begin{equation}
    \begin{split}
    e&^{-L^2|\mathbf{k'}_1+\mathbf{k}_2+\mathbf{k}_3|^2}=e^{-L^2(k^2_2+k^2_3-2k_2k_3 cos(\theta_3-\theta_2))}\\
    & \ \ e^{-L^2(k'^2_1+2k'_1k_2cos\theta_2-2k'_1k_3cos\theta_3})\\
    k&'^2= k^2_3-k^2_2+\frac{\epsilon_{\mathrm{lp}}2m}{\hbar^2}
    \end{split}
\end{equation}
and $\mathcal{H}$ the Heaviside function.
The third-order term from the out-scattering rate $\sim N^3_xN_0$ can be neglected here. The evolution of $W_{\mathrm{out}_\mathrm{c}}(T)$ as a function of the temperature is plotted in Fig.~\ref{fig:W_out_correction}.

Comparing the temperature-dependence of the scattering rates shown in Fig.~\ref{fig:W_in}, \ref{fig:W_corr}, \ref{fig:W_out1}, \ref{fig:W_out_2}, \ref{fig:W_out_correction}, we observe that the maximum value for the scattering rate is reached at a lower temperature for the in-scattering rates compared to the out-scattering rates. The letters keep increasing with $T_x$. The Maxwell-Boltzmann distribution broadens at higher temperatures, and particles have more thermal energy. This, in turn, means that particles are more likely to have higher kinetic energies and occupy higher energy states. This increased energy makes it easier for the particles to scatter out of the lower polariton state.

\end{appendix}

\nocite{*}

\bibliography{apssamp}% Produces the bibliography via BibTeX.

\end{document}